# Liquidity Risks in Lending Protocols: Evidence from Aave Protocol[1]

Xiaotong Sun[1]

Charalampos Stasinakis[2]

Georgios Sermpinis[3]

[1]University of Glasgow Business School, University of Glasgow, Gilbert Scott Building, Glasgow G12 8QQ, United Kingdom. Email: Xiaotong.Sun@glasgow.ac.uk.

[2]**Corresponding author**: University of Glasgow Business School, University of Glasgow, Gilbert Scott Building, Glasgow G12 8QQ, United Kingdom. Email: Charalampos.Stasinakis@glasgow.ac.uk,

[3]University of Glasgow Business School, University of Glasgow, Gilbert Scott Building, Glasgow G12 8QQ, United Kingdom. Email: Georgios.Sermpinis@glasgow.ac.uk.

## Abstract

Lending Protocols (LPs), as blockchain-based lending systems, allow any agents to borrow and lend cryptocurrencies. However, liquidity risks could occur, especially when salient loans are initiated by a particular group of borrowers. This paper proposes measurements of liquidity risks, focusing on both available liquidity and market concentration in LPs. By using Aave as a case study, we find that liquidity risks are highly volatile and show complex effects on Aave, and liquidity in Aave may affect across on-chain lending market. Compared to new users, regular users that repeatedly borrow cryptocurrencies may negatively affect Aave protocol, implying that user loyalty is a double-edged sword for LPs.

---

[1] The online appendix can be accessed: https://drive.google.com/file/d/1MocbHVri--8VjKQ00zix-NLQPgnE0so7/view?usp=share_link

# 1.Introduction

Financial Technology (FinTech) and its disruptive effects to traditional finance has deeply changed the financial markets (An & Rau, 2019). One of the most significant technologies is blockchain, which enables any agents to execute transactions into a ledger that is publicly observable. Some blockchains are programmable, e.g., Ethereum blockchain, and everyone can develop complicated applications based on blockchain infrastructure. Among all novel blockchain-based applications, Decentralized Finance (DeFi) has experienced rapid growth since 2019, and as of September 2021, the size of DeFi would reach $110 billion (IMF, 2021). Technically, DeFi protocols can be defined as blockchain-based financial systems, which inherit unique characteristics of blockchain, e.g., openness and transparency. The execution of transactions in DeFi does not rely on a centralized third party, e.g., central banks. Currently, DeFi can replicate most activities in traditional finance (Harvey et al., 2021; Werner et al., 2022), e.g., lending, cryptocurrency exchange, and asset management. Though we haven't witnessed systematic risks in DeFi, it is crucial to investigate risks in DeFi, which will help both investors and policy makers.

Undoubtedly, FinTech introduces convenient solutions, e.g., easier loan applications (Sun et al., 2020; Shao & Bo, 2021) and the adoption of FinTech has positive effects on the long-term growth of an economy (Kanga et al., 2021). Furthermore, empirical studies (e.g., Demir et al., 2020; Kling et al., 2020) argue that FinTech can reduce income equality. Although those advantages cannot be challenged, there are several potential risks that also cannot be ignored. One such prominent issue is FinTech-related liquidity risks. Usually, these risks are attributed to the lending market, a key segment of FinTech applications. For example, in Peer-To-Peer (P2P) lending, speculative behavior is manifested (Kanga et al., 2021). Illiquidity phenomena can also occur in some cases, e.g., when borrowers and lenders have certain trading strategies (Carison & Rose, 2019; Caglayan et al., 2020; Bechtel et al., 2022). Once illiquidity occurs, negative effects are observed, e.g., early bankruptcy (Wang et al., 2017), higher potential losses of investors (Ryu et al., 2021), and less long-term investment of banks (Choudhary & Limodio, 2022). To make matters worse, illiquidity can be contagious and liquidity risks can spread across financial markets (Aldasoro & Alves, 2018; Eross et al., 2018; Kreis & Leisen, 2018).

Stemming from this background, evaluating if liquidity risks are possible in DeFi is a very crucial task, as Lending Protocols (LPs) in DeFi are the mediums for the expansion of the lending market in FinTech. In this paper, we will focus particularly on liquidity risks in LPs, resembling banks in DeFi (Gudgeon et al., 2020a; Harvey et al., 2021). Here, liquidity risks refer to that a LP does not maintain enough available liquidity so that cannot meet the needs of withdrawing and borrowing cryptocurrencies. Without any third party, any users can easily borrow and lend cryptocurrencies by interacting with LPs. In this interaction, no documentation is required. In LPs, all activities are processed via *smart contracts*, which are rigidly coded programs. In other words, codes are the foundation of LPs, and all parameters of LPs activities are programmable. Another unique advantage of LPs is that the available liquidity is pooled in a smart contract, and anyone can examine the available liquidity and outstanding loans very easily. Different from traditional P2P lending, LP has better transparency. Currently, the main depositors contribute to most liquidity in LPs (Gudgeon et al., 2020a) and a small group of borrowers account for most loans (Saengchote, 2021). Therefore, if the main depositors withdraw their deposits successively, illiquidity problems will occur, which can then cause a market panic. Additionally, borrowers that do not repay their loan can also accelerate illiquidity problems.

The literature around liquidity risks is voluminous. Theoretically, different models of liquidity risks are developed (Bryant, 1980; Diamond & Dybvig, 1983; Rochet & Vives, 2004; Goldstein & Pauzner, 2005; Fall & Viviani, 2015). In real life, illiquidity caused unacceptable outcomes. The most influential example may be bank failures caused by financial crisis in 2008, and bank failures occurred (Hong et al., 2014). Even worse, bank defaults can cause the failure of the banking sector (Kreis & Leisen, 2018), and illiquidity can reduce banks' long-term investment, leading to negative effects on economic growth (Choudhary & Limodio, 2022). Moreover, if illiquidity occurs, it can spread across financial markets in different countries (Eross et al., 2018). Beside banks, other financial entities also suffer from illiquidity. If firms can not recognize liquidity risks, severe results, such as bankruptcy and over-leveraging, will happen (Wang et al., 2017). Badaoui et al. (2015) show that liquidity risks are also fatal in bond and CDS markets, and a series of research discuss how mutual funds' trading strategy changes because of liquidity risks (Anand et al., 2013; Collin-Dufresne & Fos, 2015; Kacperczyk & Pagnotta, 2019; Anand et al., 2021; Christoffersen et al., 2022). Since illiquidity is fatal, Allen and Gale (2004) argue that the role of regulators should be introduced, which can help to reduce liquidity risks.

However, when it comes to liquidity risks in blockchain or DeFi, the literature surprisingly remains silent. By introducing and stress-testing economic models of LPs, illiquidity is possible in some cases (Gudgeon et al., 2020b). For example, if a large price drop of collateral assets happens, LPs will be undercollateralized, and LP users may discard the risky protocols. As a result, LPs will suffer from illiquidity. Currently, most LP research is about economic models of fundamental settings and incentive mechanisms, while empirical evidence of liquidity risks is not well discussed. To fill this gap, we choose Aave, a leading LP, as a case study. Aave is founded in 2017 in Switzerland, and it raised more than $16 million in its first initial coin offering (ICO) in 2017 (Weston, 2022). Since then, Aave grows rapidly and becomes industry standards by introducing innovative functions of on-chain lending. To the best of our knowledge, this is the first paper that provides empirical evidence of liquidity risks in LP.

To achieve that, we collect information for the Aave protocol, including all borrowers, lenders and all lending-related activities from December $16^{th}$, 2019 to January $31^{st}$, 2023. By querying the intraday prices of cryptocurrencies traded in Aave, we calculate accurate available liquidity in Aave protocol. Our empirical analysis follows two stages. The first stage is to examine lending activities in Aave. We first calculate two well-adopted measurements, including liquidity and utilization, which describes a general sense of Aave protocol. We also focus on regular users, i.e., repeat borrowers and repeat depositors, who account for significant loans and deposits, respectively. Therefore, liquidity risks may occur when they collectively execute certain strategies. Based on their activities, two measurements are constructed to study the potential risks. In the second stage, we investigate the effects of potential liquidity risks in Aave protocol. Beside key metrics specific to Aave protocol, we also investigate cross-LP effects. Intuitively, liquidity can affect the underlying financial systems (Wang et al., 2017; Papanikolaou, 2018; Momtaz, 2019; Duarte et al., 2021). Therefore, factors specific to Aave protocol may change with potential liquidity risks. Besides, contagious illiquidity (Eross et al., 2018; Kreis and Leisen 2018) may also exist in LPs, so it is crucial to examine if illiquidity in Aave can affect other LPs.

The empirical results bring forward some interesting findings. First, available liquidity and utilization are highly volatile in Aave, and the spikes of utilization are very closed to 1. Furthermore, compared with new depositors and borrowers, regular Aave users (i.e., repeat depositors and repeat borrowers) contribute a large proportion of deposits and loans. By applying factor analysis, we find a complex nexus of effects of liquidity risks. For example, more liquidity and higher utilization can bring forward more revenue. However, the growth of stakeholders of Aave protocol will be refrained when there is more available liquidity. By calculating the ratio of loans initiated by repeat borrowers, we find that their effects are a double-edged sword. When repeat borrowers contribute more loans, Aave protocol benefits from better growth of market cap, and increases in the native cryptocurrency and more rapid growth of stakeholders are also a good signal. However, revenue will decrease when more loans are from repeat borrowers.

The findings are consistent with the intuitive logic around LPs. LPs rely on users to provide liquidity and initiate loans, therefore, these regular users can booster the growth of Aave by borrowing and lending. Probably, as herding exists in lending markets (Shao & Bo, 2021), these influential users can have users following them and bringing more network adoption, which is crucial in cryptofinance (Li et al., 2023; Xiong & Sockin, 2021). On the other hand, more available liquidity and better utilization are crucial for Aave. Intuitively, trading activities will be more active with sufficient liquidity, and more revenue will be generated when the liquidity is utilized. To summarize, LPs should aim to get more liquidity in the crypto market and improve the utilization of absorbed liquidity. Beside retaining regular users, it is necessary to attract new users given that regular users can cause negative effects.

More interestingly, illiquidity in Aave affects other LPs. Here, Compound[i] is chosen because it is a leading LP and main competitor of Aave. The factor analysis shows the interlinks between Aave and Compound. For example, more liquidity in Aave will decrease the growth of market cap of Compound, implying that the competition between the two LPs exists. Moreover, the return of COMP, the native cryptocurrency issued by Compound, will be lower when more liquidity goes to Aave, and the growth of stakeholders of Compound will be negatively affected. In this way, the liquidity is the position where LPs should defend, given that LPs share similar financial functions. Previous literature (e.g., Tolmach et al., 2021; von Wachter et al., 2021) focuses more on composability of DeFi protocols, which is based on the infrastructure of programmable blockchain. This paper puts emphasis on the nature of DeFi, i.e., on-chain financial systems, and the linkages among DeFi protocols, including LPs, should be investigated from the aspect of financial activities.

The remainder of this paper is organized as follows. Section 2 provides a short introduction in LPs and liquidity risks. Section 3 describes the characteristics of pool-based loans in LPs, the relevant agents and the definitions of liquidity risk measurements. Section 4 summarizes the empirical results based on the characteristics of Aave protocol and the associated regressions.

Finally, section 5 presents the conclusions, while some technical information is summarized in the Appendix section.

## 2. Lending protocols (LPs) and liquidity risks

### 2.1 Lending protocols (LPs)

Based on blockchain technology, LPs resemble banks in crypto markets, allowing their users to borrow and lend cryptocurrencies (Bartoletti et al., 2021). Any agents can lend their cryptocurrency to a LP. Similar to depositors in banks, LP depositors can earn interests by providing liquidity. Given a cryptocurrency, all deposits will be stored in a lending pool, which can be borrowed by anyone. To initiate loans, borrowers should first lock collateral, usually cryptocurrencies accepted by LPs, and both loans and interests should be repaid if a borrower aims to unlock their collateral assets. Beside borrowers and depositors, liquidators also play an important role in LPs. When borrowers fail to repay their loan, liquidators can (partly) repay the failed loan. As a result, liquidators can purchase the borrowers' collateral at a discount. Usually, the process is defined as liquidation. Compared to traditional bank lending, a pivotal difference of LPs is that key parameters of loans are not decided by third parties. For example, LPs apply different mathematical models to determine the interest rates. Other suggested changes, e.g., new acceptable collateral, will be usually jointly decided by LP users via voting. For more details, we refer readers to summary research presented by Gudgeon et al. (2020a) and Werner et al. (2022).

#### 2.1.1 Aave protocol

Currently, diversified LPs co-exist in DeFi, while several mainstream LPs account for the most lending activities. One of the most widely adopted LPs is *Aave*[ii], and Aave provides service on multiple blockchains. In this paper, we focus on Aave on Ethereum blockchain. The first version of Aave protocol, i.e., Aave V1, was deployed to the Ethereum mainnet in January 2020. Aave updated to its second version, i.e., Aave V2, in December 2020. As of January 31$^{st}$, 2023, the total locked value (in USD) in Aave is more than $4.5 billion.

Beside the significant market capitalization, Aave expands its influence by introducing new features. First, Aave allows any users to create lending pools. Theoretically, Aave users can initiate loans in any cryptocurrencies that they prefer. This is unlike any relevant flexibility appearing in traditional banking lending. Second, Aave was the first LP to introduce '*flash* *loans*'. To put it simply, flash loans do not require any collateral, since the loan will be borrowed and repaid in an atomic transaction group (Qin et al., 2021). More details about utilization of flash loans can be found in Gudgeon et al. (2020a), Wang et al. (2021) and Qin et al. (2021). This function allows more possibility of on-chain transactions. Aave introduces also several innovative features in V2, such as swapping collateral assets and repaying debts with collateral assets (Aave, 2021). These innovative technical features not only increase the adoption of Aave, but also make it the main industry standard.

## 2.2 Liquidity risks

Liquidity risks can relate to several subfields in financial studies. In the banking literature, liquidity attracts much attention because liquidity creation is one of the core functions of banks (Diamond, 1984; Berger & Bouwman, 2009). Theoretically, Brynat (1980) and Diamond and Dybvig (1983) first introduce models of illiquidity in banks (e.g., 'bank run' models). Some recent studies (Fall & Viviani, 2015) also contribute to theoretical models of liquidity risks. To address this problem, banks should maintain sufficient funds to meet the withdrawal needs of their customers. Furthermore, liquidity also has risk transfer functions and allow banks to make profits by financing risky loans (Holmström & Tirole, 1998), so liquidity is the core of financial systems in banks. In this paper, liquidity risks refer to that a LP does not have sufficient available liquidity so that probably fails to meet the needs of withdrawing and borrowing cryptocurrencies.

The most widely discussed example of liquidity risks is global financial crisis in 2008. Liquidity risks resulted in bank failures in 2009 and 2010 (Hong et al., 2014), which severely affected global economy. Eross et al. (2018) and Kreis and Leisen (2018) also address the infectivity of liquidity risks across countries. To better understand how banks should act when illiquidity is possible, Morris and Shin (2004) and Goldstein and Pauzner (2005) calculate the possibility of the bank to survive when liquidity risks occur. They find that private information is crucial during illiquidity crisis, and banks can reduce liquidity risks by constructing demand-deposit contracts. Beside global crisis, short-term liquidity risks can also negatively impact bank stability (Wagner, 2007; Acharya & Naqvi, 2012). Generally, it is accepted that banks that have better liquidity usually have better market performance. This is shown, for example, by Berger and Bouwman (2009) who argue that liquidity is positively related to the bank value, by examining a sample of the US banks from the period 1993-2003.

As for the reasons of liquidity risks, the literature provides several explanations. One main explanation is deposit volatility which can turn have in turn an illiquidity effect. Choudhary and Limodio (2022) show that such risks can even have long-term negative effects. The structure of interest rates is also tied with liquidity risks, especially in lending markets (Bechtel et al., 2022). Ownership networks are also crucial in liquidity risks from a banking perspective. Gao et al. (2022) show that there exists a relationship between social networks and liquidity risks, which may be related to private information. Their study focuses on loan-level data in China. Aldasoro and Alves (2018) and Berndsen et al. (2018) also point out the importance of networks in financial stability. In some cases, financial networks can encourage risk-taking behavior (Silva et al., 2016). Finally, the quality of collateral is crucial. Usually, low-quality collateral is a dangerous signal when discussing liquidity risks (Macchiavelli & Pettit, 2020).

**3. Pool-based loans in LPs: Depositors, borrowers, liquidators and associated liquidity risks.**

In this section, we will present a model to better describe activities in LPs. Our model features three types of agents, i.e., depositors, borrowers, and liquidators; and five types of activities, i.e., borrow, repay, deposit, withdraw, liquidation. Depositors lend cryptocurrencies to LPs, while borrowers borrow cryptocurrencies by locking collateral assets. Once a borrower fails to repay his loan, or his debt is undercollateralized, a liquidator can partly repay the loan and purchase collateral assets at a discounted price. All validated activities are publicly observable by all agents. Figure 1 illustrate how borrowers and depositors interact with LPs. If a LP support loan of cryptocurrency $A$, a liquidity pool will be generated, where users can deposit or borrow cryptocurrency $A$.

[Figure 1 here]

**3.1 Depositors**

Depositors will receive an amount of claim after transferring cryptocurrencies to a LP. The claim is a cryptocurrency minted by a LP and it is a proof of deposits. For depositors, the amount of claim received will correspond to the amount of deposits. The claim will be redeemable for a value of the same cryptocurrency type of the original deposit. So, when depositors want to withdraw their deposits, they need to transfer claim to LPs.

Given a cryptocurrency $A$, we assume there exist $N$ depositors, indexed by $i \in \{1,2,...,N\}$, whose deposits on date $t$ are $d_{A,i,t}, i \in \{1,2,...,N\}$ and $t \in \{1,2,...,T\}$. For depositor $i$, his withdrawn deposits on date $t$ are $w_{A,i,t}, i \in \{1,2,...,N\}$ and $t \in \{1,2,...,T\}$.

So, the supply of cryptocurrency $A$ from depositor $i$ on date $t$:

$$supply_{A,i,t} = \sum_{s=1}^{t}(d_{A,i,s} - w_{A,i,s}) \quad (1)$$

And the outstanding deposits of cryptocurrency $A$ on date $t$:

$$outstanding\ deposit_{A,t} = \sum_{i=1}^{N} supply_{i,t} \quad (2)$$

### 3.2 Borrowers

Borrowers can initiate loans from a LP if only they lock enough collateral. Usually, overcollateralization is required (Bartoletti et al., 2021), meaning that the value of debt is lower than the value of collateral. Collateral can be cryptocurrencies supported by LPs and will be locked in the loan duration.

Given a cryptocurrency $A$, we assume there exist $M$ borrowers, indexed by $j \in \{1,2,...,M\}$, whose loans on date $t$ are $b_{j,t}, j \in \{1,2,...,M\}$ and $t \in \{1,2,...,T\}$. For borrower $j$, his repaid loans on date $t$ are $r_{j,t}, j \in \{1,2,...,M\}$ and $t \in \{1,2,...,T\}$.

So, the demand of cryptocurrency $A$ from borrower $j$ on date $t$:

$$demand_{A,j,t} = \sum_{s=1}^{t}(b_{A,j,s} - r_{A,j,s}) \quad (3)$$

### 3.3 Liquidators

Once borrowers fail to repay their loans, or debt is undercollateralized, liquidators can (partly) repay the loans to acquire a discount amount of collateral (see Figure 2). In LPs, the process is called liquidation (Kao et al., 2020). The liquidation thresholds vary between asset markets across different protocols (Gudgeon et al., 2020a).

[Figure 2 here]

Given a cryptocurrency $A$, we assume there exist $L$ liquidators, indexed by $k \in \{1,2,\ldots,L\}$, whose repaid loans on date $t$ are $l_{k,t}, k \in \{1,2,\ldots,L\}$ and $t \in \{1,2,\ldots,T\}$. On date $t$, the total loans of cryptocurrency $A$ repaid by all liquidators:

$$l_{A,t} = \sum_{s=1}^{t} \sum_{k=1}^{L} l_{A,k,s} \quad (4)$$

Consequently, loans will be paid by borrowers and liquidators, the outstanding debt of cryptocurrency $A$ on date $t$:

$$outstanding\ debt_{A,t} = \sum_{j=1}^{M} demand_{A,j,t} - l_{A,t} \quad (5)$$

## 3.4 Available liquidity and utilization in LPs

For a cryptocurrency in LPs, liquidity means that the total supply is more than the total demand. We can calculate the available liquidity of cryptocurrency $A$ on date $t$:

$$liquidity_{A,t} = outstanding\ deposit_{A,t} - outstanding\ debt_{A,t} \quad (6)$$

Then, the USD value of available liquidity of all cryptocurrencies in a LP on date $t$ can be calculated as

$$liquidity_t = \sum_{A} liquidity_{A,t} \times p_{A,t} \quad (7)$$

Where $p_{A,t}$ is the price of cryptocurrency $A$ on date $t$.

The utilization of cryptocurrency $A$ on date $t$:

$$utilization_{A,t} = \frac{outstanding\ debt_{A,t}}{outstanding\ deposit_{A,t}} \quad (8)$$

The utilization of a LP on date $t$ can be calculated as

$$utilization_t = \frac{\sum_{A} outstanding\ debt_{A,t} \times p_{A,t}}{\sum_{A} outstanding\ deposit_{A,t} \times p_{A,t}} \quad (9)$$

Where $p_{A,t}$ is the price of cryptocurrency $A$ on date $t$.

Traditional examples of measurements of liquidity risks in banking are given by Holmström and Tirole (1998), Berger and Bouwman (2009) and Fall and Viviani (2015). However, these comprehensive measures rely on balance sheet information and how loans are classified. This is not suitable for LPs. We choose available liquidity and utilization as two simple and intuitive measurements. In practice, liquidity risks can be triggered by successive withdrawals, especially when the depositors with large deposits decide to leave (Alethio, 2020). In this case, available liquidity in LPs will be low, while utilization should be close to one.

## 4. Empirical Results

### 4.1 Measurements of liquidity risks

In this paper, we focus on Aave on Ethereum blockchain. *Aave: LendingPool V1* [iii] and *Aave: LendingPool V2* [iv] are the main components of Aave protocol, and these two contracts document events related to deposits and loans in Aave V1 and V2, respectively. Utilizing *Dune.xyz*, we query all transactions in Aave from December 16th, 2019 to January 31st, 2023. For every transaction, we retrieve the real-time prices (in USD) of the borrowed or deposited cryptocurrency, therefore, the statistics, such as the daily volume (in USD) of loans and deposits, can be precisely calculated. Table 1 summarizes the details of loans and deposits in Aave. Overall, we do not observe many borrowers or depositors, though the means of daily volume of loans and deposits are more than $20 million. We also consider regular borrowers and depositors in Aave (here after, *repeat borrower* and *repeat depositor*). Comparing to new borrowers and depositors, on average, regular users usually contribute more loans and deposits to Aave protocol.

[Insert Table 1 here]

Figure 3 divulgate more information about liquidity in Aave protocol. Overall, outstanding debt and deposit (in USD) experienced rapid growth around June 2021 and have shown volatility since then. Noticeably, outstanding debt was extremely closed to deposit at some point, implying potential illiquidity. Then, we compute two measurements, i.e., *liquidity* and *utilization*. Daily available liquidity is highly volatility after December 2021, while the spikes of utilization are more than 0.8. All these signals suggest that Aave is not immune from liquidity risks.

[Insert Figure 3 here]

To better investigate the influence of repeat borrowers and depositors, we first compute their activities in Aave protocol. Figures 4 and 5 show that, though Aave keeps attracting new depositors and borrowers[v], repeat depositors and borrowers account for most volume (in USD) of deposits and loans, respectively. Then, we construct two ratios related to repeat borrowers and repeat depositors, namely *repeat loan ratio* and *repeat deposit ratio*. *Repeat loan ratio* is the proportion of daily volume (in USD) of loans contributed by repeat borrowers, while *repeat deposit ratio* equals to the proportion of daily volume (in USD) of deposits contributed by repeat depositors. When the two ratios are higher, illiquidity is more likely if repeat borrowers or repeat depositors collectively initiate loans and withdraw deposits. Overall, these two measurements are very volatile and usually higher than 0.8, implying that repeat users are dominant players in Aave. Table 2 presents the descriptive statistics of four measurements of liquidity risks.

**[Insert Figures 4 - 6 here]**

**[Insert Table 2 here]**

### 4.2 Effects of liquidity risks on Aave protocol

Naturally, we are interested in how liquidity risks affect Aave protocol. To achieve that, we construct a series of factors specific to Aave protocol, and the original datasets are available on *Dune.xyz* and *tokenterminal.com.* Table 3 gives brief introduction to factors related to Aave, though some of them are not included in the regression models[vi].

**[Insert Table 3 here]**

Then, we estimate the regression models below:

$$Protocol_{i,t} = \beta_0 + \beta_1 Risk_t + \beta_2 \Delta Outstanding\ loan_{i,t} + \beta_3 \Delta Outstanding\ deposit_{i,t}$$
$$+ \beta_4 Deposit\ vol\ usd_{i,t} + \beta_5 Loan\ vol\ usd_{i,t} + \beta_6 Liquidation\ usd_{i,t}$$
$$+ \beta_7 \Delta Active\ user_{i,t} + \beta_8 \Delta Developer_{i,t} + \varepsilon_{i,t} (10)$$

Where:

- $Risk = \{Liquidity, Utilization, Repeat\ deposit\ ratio, Repeat\ loan\ ratio\}$
- $Protocol = \{\Delta MktC\_F, \Delta MktC\_C, Revenue, \Delta TVL, \Delta AAVE, \Delta AAVE\ holder\}$
- $i = Aave\ protocol$

The aim in this regression setup is to used dependent variables able to describe the status of Aave protocol. Market cap and total value locked (TVL) are two widely adopted metrics of project performance. Usually, higher market cap and TVL are positive signals. We also consider daily revenue (in USD), which also reflects on the state of Aave. Besides, two variables related to *AAVE token* are studied. AAVE token, also a cryptocurrency, is the governance token of Aave protocol, and AAVE holders can participate in autonomous governance of Aave protocol. In a sense, AAVE resembles stocks issued by corporations, while AAVE holders play a similar role with shareholders. Therefore, the return of AAVE and the number of AAVE holders show the market expectations and the evaluation on Aave protocol.

Beside the liquidity risk measurement, we have other several dependent variables that help to explain the changes of Aave-specific factors. Given that Aave centers on on-chain lending, both outstanding loans and deposits are crucial determinants of market performance. To capture the intraday activities, we consider daily volume of loans and deposits, and the value (in USD) of liquidated collateral assets. Besides, the number of active users should be included because the importance of network adoption. In the context of crypto markets, Li et al. (2023) and Xiong and Sockin (2021) contend that better network adoption can booster valuation and long-term development of on-chain financial systems. Finally, we include the number of developers, given that these people can influence the technical updates of Aave protocol.

The results of these regressions are summarized in the following table. First, *liquidity* and *utilization* can increase revenue of Aave protocol, while *repeat loan ratio* shows negative effects. When the liquidity is adequate or the deposited cryptocurrencies are sufficiently utilized, i.e., borrowed, Aave protocol consequently earn more revenue, since the revenue is relevant to lending activities. But high *repeat loan ratio* can be a concern. If more loans are initiated by repeat borrowers, Aave protocol suffers from lower revenue, implying the necessity of attracting new users. This point is consistent with the arguments of Li et al. (2023) and Xiong and Sockin (2021), where network adoption is crucial for on-chain financial applications.

[Insert Table 4 here]

*Repeat loan ratio* also show other influences on Aave protocol. Increases in *repeat loan ratio* can contribute to the growth of market cap, the return of AAVE, and the growth of AAVE holders. The findings make us think about the role of repeat borrowers. If a DeFi aims for a large market share and numerous stakeholders, they should make efforts to retain regular users, and these regular users' activities can thereby increase the return of the native governance cryptocurrencies. In this way, both new users and regular users are pivotal for DeFi, which is in line with common sense. The positive relationship between AAVE return and *repeat loan ratio* also provides new insights of asset returns. Momatz (2019) and Zhang and Li (2021) provide empirical evidence that liquidity is a predictor of cryptocurrency return, however, the previous literature usually ignores the feature of cryptocurrency traders (e.g., if they are regular borrowers of LPs). In the context of corporate finance, after observing companies' liquidity risks, investors can have certain trading strategies and thereby cause negative dynamics of stock returns (Roogi & Giannozzi, 2015).

**4.3 Cross-LP effects of liquidity risks**

LPs cannot be seen through an isolated lens on blockchain. Benefiting from blockchain infrastructure, DeFi protocols can be built on and interact with other protocols. This element of composability of DeFi (Tolmach et al., 2021; von Wachter et al., 2021) is quite important for the validity of each protocol. For that reason, we should also be investigating illiquidity contagion also in LPs, as in traditional financial markets. To examine such cross-LP effects, we choose Compound, which is a leading LP and strong competitor of Aave protocol. The definitions of Compound factors are similar to Aave factors (Appendix 2 presents the definitions of Compound factors, and OA.2 explains selection of dependent variables in regression models).

To that end, we estimate the following regression models:

$$Protocol_{i,t} = \beta_0 + \beta_1 Risk_t + \beta_2 \Delta Outstanding\ loan_{i,t} + \beta_3 \Delta Outstanding\ deposit_{i,t} + \beta_4 Loan\ vol\ usd_{i,t} + \beta_5 Liquidation\ usd_{i,t} + \beta_6 Active\ user_{i,t} + \beta_7 \Delta Developer_{i,t} + \varepsilon_{i,t} (11)$$

Where:

- $Risk = \{Liquidity, Utilization, Repeat\ deposit\ ratio, Repeat\ loan\ ratio\}$
- $Protocol = \{\Delta MktC\_F, \Delta MktC\_C, \Delta Revenue, \Delta TVL, \Delta COMP, \Delta COMP\ holder\}$
- $i = Compound\ protocol$

Following similar logic in factor selection for Compound as in Aave, we present the results of these regression in the following tables.

**[Insert Table 5 here]**

Interestingly, liquidity in Aave shows relation to Compound protocol. More liquidity in Aave will lower the growth of market cap of Compound, and the return of COMP token and the number of COMP holders will decrease as well. Here, COMP token, also a cryptocurrency, is the governance token of Compound protocol, and COMP holders resembles shareholders in corporations. Our findings contend that liquidity risks in one LP may result in worse performance of other LPs. The finding is intuitive because Aave and Compound are substitutes for each other. When more liquidity is absorbed by Aave, it is normal that Compound is negatively affected. For the case of other three measurement of liquidity risks in Aave, they do not show statistically significant results. Based on the empirical findings, we expect that more empirical work should be done for better studying if liquidity risks are infectious in DeFi ecosystem, especially when DeFi hacks happen.

**4.4 The influences of technical updates**

On December 3$^{rd}$, 2020, Aave protocol V2 (hereafter, Aave V2) was launched, and innovative features, e.g., flash loans, were introduced by Aave V2[vii]. One potential issue is that the empirical findings may lack robustness if we incorporate the Aave V2 update. To address this issue, we estimate the following regression models:

$$Protocol_{i,t} = \beta_0 + \beta_1 Risk_t + \beta_2 V2_{i,t} + \beta_3 \Delta Outstanding\ loan_{i,t}$$
$$+ \beta_4 \Delta Outstanding\ deposit_{i,t} + \beta_5 Deposit\ vol\ usd_{i,t} + \beta_6 Loan\ vol\ usd_{i,t}$$
$$+ \beta_7 Liquidation\ usd_{i,t} + \beta_8 \Delta Active\ user_{i,t} + \beta_9 \Delta Developer_{i,t} + \varepsilon_{i,t} (12)$$

Where:

- $Risk = \{Liquidity, Utilization, Repeat\ deposit\ ratio, Repeat\ loan\ ratio\}$
- $Protocol = \{\Delta MktC\_F, \Delta MktC\_C, Revenue, \Delta TVL, \Delta AAVE, \Delta AAVE\ holder\}$
- $V2_t = \begin{cases} 0, t < 2020-12-03 \\ 1, t \geq 2020-12-03 \end{cases}$
- $i = Aave\ protocol$

Compared with the model (10), a dummy *V2* is included to study the influences of update of Aave V2. Table 6 demonstrates that the empirical results remain valid even after introducing the dummy variable.

**[Insert Table 6 here]**

For the case of Compound, we also consider its technical updates. Here, we consider Compound V3[viii], which was deployed on August 25th, 2022, and a dummy variable *V3* is constructed. Then, we estimate the following regression models:

$$Protocol_{i,t} = \beta_0 + \beta_1 Risk_t + \beta_2 V3_{i,t} + \beta_3 \Delta Outstanding\ loan_{i,t}$$
$$+ \beta_4 \Delta Outstanding\ deposit_{i,t} + \beta_5 Loan\ vol\ usd_{i,t} + \beta_6 Liquidation\ usd_{i,t}$$
$$+ \beta_7 Active\ user_{i,t} + \beta_8 \Delta Developer_{i,t} + \varepsilon_{i,t} (13)$$

Where:

- $Risk = \{Liquidity, Utilization, Repeat\ deposit\ ratio, Repeat\ loan\ ratio\}$
- $Protocol = \{\Delta MktC\_F, \Delta MktC\_C, \Delta Revenue, \Delta TVL, \Delta COMP, \Delta COMP\ holder\}$
- $V3_t = \begin{cases} 0, t < 2022-08-25 \\ 1, t \geq 2022-08-25 \end{cases}$
- $i = Compound\ protocol$

The following table presents the empirical results. Again, introducing the dummy variable does not disturb the results in section 4.3.

[Insert Table 7 here]

**4.5 The shocks of DeFi hacks**

A bedevilled issue of DeFi is hacking. Since codes are the foundation of DeFi, DeFi is vulnerable to hacking and bugs. The table below lists the most severe DeFi hacks happened from 2020 to 2022. In the banking sector, individual bank defaults are infectious and can cause the failure of the banking sector (Kreis & Leisen, 2018). Though Compound and Aave were not the victims in the chosen hacks, we expect to study if DeFi hacks can affect these two LPs.

[Insert Table 8 here]

To that end, we construct a dummy variable *hack*, and *hack* equals to 1 when a hack happened. Given that the negative influences of hacking usually continues for several days, we let *hack* equal to 1 on the following 6 days of a hack. Then, we estimate the regression models below:

$$Protocol_{i,t} = \beta_0 + \beta_1 Risk_t + \beta_2 Hack_{i,t} + \beta_3 \Delta Outstanding\ loan_{i,t}$$
$$+ \beta_4 \Delta Outstanding\ deposit_{i,t} + \beta_5 Deposit\ vol\ usd_{i,t} + \beta_6 Loan\ vol\ usd_{i,t}$$
$$+ \beta_7 Liquidation\ usd_{i,t} + \beta_8 \Delta Active\ user_{i,t} + \beta_9 \Delta Developer_{i,t} + \varepsilon_{i,t} (14)$$

Where:

- $Risk = \{Liquidity, Utilization, Repeat\ deposit\ ratio, Repeat\ loan\ ratio\}$
- $Protocol = \{\Delta MktC\_F, \Delta MktC\_C, Revenue, \Delta TVL, \Delta AAVE, \Delta AAVE\ holder\}$
- $Hack_t = \begin{cases} 1, a\ hack\ happend\ between\ day\ t\ and\ day\ t-6 \\ 0, otherelse \end{cases}$
- $i = Aave\ protocol$

Table 9 presents the results, which are consistent with results in section 4.2. Furthermore, *hack* shows positive effects on revenue of Aave protocol and negative effects on the growth of AAVE holders. Given that the chosen hacks did not attack Aave protocol, more users may circumvent the victims and choose Aave protocol. As a result, the revenue of Aave protocol increased in the short period after a hack. The decreased growth of AAVE holders seem to be more mysterious. A possible explanation is that DeFi users are not willing to be stakeholders of Aave protocol after witnessing a hack, but the explanation may be contrary to the higher revenue of Aave after a hack.

**[Insert Table 9 here]**

For the case of Compound protocol, we estimate the following regression models:

$$Protocol_{i,t} = \beta_0 + \beta_1 Risk_t + \beta_2 Hack_{i,t} + \beta_3 \Delta Outstanding\ loan_{i,t}$$
$$+ \beta_4 \Delta Outstanding\ deposit_{i,t} + \beta_5 Loan\ vol\ usd_{i,t} + \beta_6 Liquidation\ usd_{i,t}$$
$$+ \beta_7 Active\ user_{i,t} + \beta_8 \Delta Developer_{i,t} + \varepsilon_{i,t} (15)$$

Where:

- $Risk = \{Liquidity, Utilization, Repeat\ deposit\ ratio, Repeat\ loan\ ratio\}$
- $Protocol = \{\Delta MktC\_F, \Delta MktC\_C, \Delta Revenue, \Delta TVL, \Delta COMP, \Delta COMP\ holder\}$
- $Hack_t = \begin{cases} 1, a\ hack\ happend\ between\ day\ t\ and\ day\ t-6 \\ 0, otherelse \end{cases}$
- $i = Compound\ protocol$

The results are presented in the following table, which are consistent with the findings in section 4.2. Noticeably, *hack* does not show influences on Compound protocol, implying that Compound can better fend off financial risks caused by hacking.

**[Insert Table 10 here]**

**4.6 Mainstream cryptocurrencies**

Among all cryptocurrencies traded on Aave, there are some cryptocurrencies that are more popular and account for most borrowing and lending activities. In this subsection, we select 16 most frequently traded cryptocurrencies (hereafter, mainstream cryptocurrencies) and construct a subset to examine if our empirical results still hold. The following table gives a brief introduction to the selected cryptocurrencies, and more details about the relevant borrowing and lending activities are given in appendix A.3.

**[Insert Table 11 here]**

Using the subset, we first compute the measurements of liquidity risks for mainstream cryptocurrencies, and the descriptive statistics are given in the following table. Overall, the statistics based on mainstream cryptocurrencies are closed to the statistics based on all cryptocurrencies traded on Aave. Therefore, we expect to observe empirical findings similar to sections 4.2 and 4.3.

**[Insert Table 12 here]**

In this subsection, we also consider the influences of Aave V2, Compound V3, and DeFi hacks, and the regression models below are estimated:

$$Protocol_{i,t} = \beta_0 + \beta_1 Risk_t + \beta_2 V2_{i,t} + \beta_3 Hack_{i,t} + \beta_4 \Delta Outstanding\ loan_{i,t}$$
$$+ \beta_5 \Delta Outstanding\ deposit_{i,t} + \beta_6 Deposit\ vol\ usd_{i,t} + \beta_7 Loan\ vol\ usd_{i,t}$$
$$+ \beta_8 Liquidation\ usd_{i,t} + \beta_9 \Delta Active\ user_{i,t} + \beta_{10} \Delta Developer_{i,t} + \varepsilon_{i,t} (16)$$

Where:

- $Risk = \{Liquidity, Utilization, Repeat\ deposit\ ratio, Repeat\ loan\ ratio\}$
- $Protocol = \{\Delta MktC\_F, \Delta MktC\_C, Revenue, \Delta TVL, \Delta AAVE, \Delta AAVE\ holder\}$
- $V2_t = \begin{cases} 0, t < 2020 - 12 - 03 \\ 1, t \geq 2020 - 12 - 03 \end{cases}$
- $Hack_t = \begin{cases} 1, a\ hack\ happend\ between\ day\ t\ and\ day\ t - 6 \\ 0, otherelse \end{cases}$
- $i = Aave\ protocol$

The following table summarizes the results for liquidity risks caused by mainstream cryptocurrencies. Not surprisingly, the results are consistent with our previous findings in section 4.2.

**[Insert Table 13 here]**

To study how liquidity risks caused by mainstream cryptocurrencies affect Compound protocol, we estimate the following regression models:

$$Protocol_{i,t} = \beta_0 + \beta_1 Risk_t + \beta_2 V3_{i,t} + \beta_3 Hack_{i,t} + \beta_4 \Delta Outstanding\ loan_{i,t}$$
$$+ \beta_5 \Delta Outstanding\ deposit_{i,t} + \beta_6 Loan\ vol\ usd_{i,t} + \beta_7 Liquidation\ usd_{i,t}$$
$$+ \beta_8 Active\ user_{i,t} + \beta_9 \Delta Developer_{i,t} + \varepsilon_{i,t} (17)$$

Where:

- $Risk = \{Liquidity, Utilization, Repeat\ deposit\ ratio, Repeat\ loan\ ratio\}$
- $Protocol = \{\Delta MktC\_F, \Delta MktC\_C, \Delta Revenue, \Delta TVL, \Delta COMP, \Delta COMP\ holder\}$
- $V3_t = \begin{cases} 0, t < 2022-08-25 \\ 1, t \geq 2022-08-25 \end{cases}$
- $Hack_t = \begin{cases} 1, a\ hack\ happend\ between\ day\ t\ and\ day\ t-6 \\ 0, otherelse \end{cases}$
- $i = Compound\ protocol$

The following table presents the results, which are mostly consistent with the findings in section 4.3. For example, *liquidity* shows negative effects on the growth of market cap, the return of COMP, and the growth of COMP holders.

**5. Conclusion**

An important question always posed towards the community of FinTech practitioners and entrepreneurs is whether FinTech can be immune from risks observed in traditional financial markets. This paper is attempting to shed light in this question from the lens of one of the controversial FinTech segments, LPs. Particularly, we focus on liquidity risks and we examine if liquidity risks exist in LPs, which are blockchain-based lending systems. Several studies discuss how banks suffer from illiquidity (e.g., Brynat, 1980; Diamond & Dybvig, 1983; Allen & Santomero, 1997; Allen & Gale, 2004; Ryu et al., 2021). Recent studies show that P2P lending also faces illiquidity (Käfer, 2017), and other problems may exist, e.g., risks caused by borrowers with lower credit grade (Emekter et al., 2014), asymmetric information faced by P2P participants (Shao & Bo, 2021), and regional discrimination (Wang et al., 2020). However, LPs, as the emerging on-chain lending systems have not been scrutinized in that regard.

To study liquidity risks in LPs, we focus on the Aave protocol and the relevant lending activities. Not surprisingly, the available liquidity in Aave can be very close to zero sometimes, implying that illiquidity is possible. After applying a series of factor analysis, we find that the available liquidity can contribute to more revenue of Aave protocol but increases in liquidity can negatively affected the growth of AAVE holders, who are stakeholders of Aave protocol. The findings first extend the research on the relationship between bank liquidity and profitability. For example, Molyneux and Thornton (1992), among others, show that liquidity can negatively influence profitability. However, the opposite arguments, such as Bourke (1989) and Al-Matari (2021), are also presented. This paper further contributes to the debates, and the findings shows the complicated role of available liquidity in LPs.

From our descriptive analysis of the Aave data, we notice that a small group of Aave users are regular borrowers and depositors, who provide significant proportion of stablecoin liquidity and initiate large loans. In stock markets, investors can react to liquidity risks of companies and then cause price dynamics of stocks (Roggi & Giannozzi, 2015). Moreover, investors can make trade-offs among risks. For example, shareholders of banks may be more intolerant to interest rate risks so that liquidity risks are relatively 'acceptable' (Chen & Lin, 2016). In LPs, repeat borrowers and repeat depositors may affect the protocol via their activities. Therefore, we make efforts to examine the influences the two categories of users in Aave by constructing two measurements for potential liquidity risks caused by them.

Our empirical findings show that the repeat borrowers are a double-edged sword. For example, when more loans are issued by these users, in other words, potential risks caused by these users are higher, Aave can have better growth of market cap, and the return of its governance token, namely AAVE, will increase. This also leads to quicker growth of AAVE holders. In other words, these users can booster Aave. On the other hand, the Aave protocol may suffer from their behavior. When repeat borrowers account for more loans, the revenue of Aave protocol will be negatively affected. For LPs, retaining regular users is a primary goal, given that market cap and better adoption of the native governance token are key metrics of evaluating on-chain financial systems. However, the decreasing revenue may undermine the long-term development of LPs.

Our results are also extended in the potential contagion of illiquidity across LPs, which is a logical step as in traditional financial markets. To capture these potential cross-LPs effects of liquidity risks, we choose Compound, which is a leading LP and main competitor of Aave. The findings indicate that liquidity risks bring forward interlinks between these two LPs. For example, more liquidity absorbed by Aave will decelerate the growth of market cap and the growth of COMP holders, implying that users tend to adopt LPs with more sufficient liquidity.

Literature of banking studies (e.g., Kreis & Leisen, 2018) addresses that illiquidity risks of individual banks can cause the crisis of the banking sector. Though we have not witnessed severe defaults of Aave or Compound, the interlinks via absorbed liquidity imply that default linkages of LPs are possible.

Although our findings appear conceptually and empirically robust, they should be interpreted with their limitations in mind. First, we do not show the identity of repeat borrowers or repeat depositors, which can help us understand their trading strategy. However, in blockchain, anonymity is another unique characteristic. Unless these users are willing to announce their identity, we will hardly ever know who they are. Second, potential liquidity risks can be affected by social networking. Assuming that potential collusion could exist among a group of malicious liquidity providers, there can be exposure to higher risk of illiquidity attacks. This possibility of successful illiquidity attacks is not captured by our constructed liquidity risk measurements. To better predict liquidity risks, we may need private information, such as the relationship of LP users and their strategy in different market conditions, which at this stage is extremely difficult to secure. Finally, better measurements of liquidity risks in LPs may be a good topic for further study. In traditional finance, how to measure liquidity risks is a controversial topic. Lou and Sadka (2011) argue that liquidity level and liquidity risks are not the same thing. In this paper, we do not investigate the overall liquidity level of Aave protocol, given that there are various cryptocurrencies traded on Aave. A possible choice is expected default probability (EDF), and Covitz and Downing (2007) and Erkens et al. (2012) show how to apply the measurements. We expect to see new measurements that can better describe liquidity risks in LPs.

**Tables**

**Table 1. Descriptive statistics of details of loans and deposits in Aave**

| Panel A: Loan details | | | | | |
|---|---|---|---|---|---|
| **Variable** | **Mean** | **Median** | **Maximum** | **Minimum** | **Std** |
| **Borrower** | 180.26 | 160 | 2202 | 3 | 137.29 |
| **Loan vol usd** | 74402450.33 | 44690200.14 | 1312822432.44 | 615.84 | 111355283.92 |
| **Loan cnt** | 257.92 | 223 | 2265 | 9 | 182.73 |
| **New borrower** | 49.20 | 34 | 2113 | 1 | 93.23 |
| **New loan vol usd** | 10839280.04 | 2152363.60 | 452990005.67 | 27.91 | 35010964.38 |
| **New loan cnt** | 58.94 | 42 | 2143 | 1 | 96.74 |
| **Avg loan usd** | 279693.06 | 171118.75 | 6132660.20 | 23.69 | 380383.67 |
| **Outstanding loan** | 2445509004.02 | 1986428589.00 | 8583137562.00 | 0.00 | 2407001377.99 |
| **Liquidation usd** | 1108629.00 | 8778.63 | 182148137.30 | 0 | 7192831.19 |
| **Repeat borrower** | 131.07 | 110 | 548 | 0 | 90.83 |
| **Repeat loan vol usd** | 63563170.29 | 38763466.99 | 1293039942.36 | 0 | 94246681.88 |
| **Repeat loan cnt** | 198.98 | 168 | 903 | 0 | 137.70 |
| **Panel B: Deposit details** | | | | | |
| **Variable** | **Mean** | **Median** | **Maximum** | **Minimum** | **Std** |
| **Depositor** | 240.13 | 210 | 2315 | 1 | 161.27 |
| **Deposit vol usd** | 244207804.95 | 135861456.56 | 4353999722.71 | 3.01 | 377703443.59 |
| **Deposit cnt** | 478.64 | 401 | 2643 | 1 | 321.52 |
| **New depositor** | 94.78 | 73 | 2117 | 1 | 104.30 |
| **New deposit vol usd** | 32321743.93 | 7459112.17 | 3238493811.48 | 3.01 | 134629694.90 |
| **New deposit cnt** | 121.69 | 95 | 2194 | 1 | 119.78 |
| **Avg deposit usd** | 555875.97 | 281998.74 | 11549070.88 | 3.01 | 1026335.57 |
| **Outstanding deposit** | 5953130710.29 | 7604941310.00 | 16367590242.00 | 0 | 5015574904.39 |
| **Repeat depositor** | 145.34 | 123 | 619 | 0 | 86.82 |
| **Repeat deposit vol usd** | 211886061.02 | 117195291.80 | 4236981879.53 | 0 | 326516103.30 |
| **Repeat deposit cnt** | 356.95 | 296 | 1985 | 0 | 247.94 |

Note: This table reports details of loans and deposits in Aave v1 and v2 (from Dec 16th, 2019 to Jan 31st, 2023). In this table, we consider all tokens traded in Aave protocol. The definitions of variables are presented in table A.1.

**Table 2. Measurements of liquidity risks**

| | **Liquidity** | **Utilization** | **Repeat deposit ratio** | **Repeat loan ratio** |
|---|---|---|---|---|
| **Mean** | 3667536822.32 | 0.36 | 0.86 | 0.86 |
| **Median** | 3924483726.00 | 0.31 | 0.92 | 0.93 |
| **Maximum** | 10122920707.00 | 0.84 | 1.00 | 1.00 |
| **Minimum** | 0.00 | 0 | 0.00 | 0.00 |
| **Std** | 3090386938.75 | 0.16 | 0.16 | 0.16 |

Note: This table reports descriptive statistics of liquidity risk measurements based on the datasets for Aave V1 and V2 lending pool (from Dec 16th, 2019 to Jan 31st, 2023).

**Table 3. Aave protocol-specific factors**

| Factor | Definition |
|---|---|
| **MktC_F** | Market cap (in USD) based on the maximum supply of tokens |
| **MktC_C** | Market cap (in USD) based on the circulating supply of tokens |
| **AAVE** | Daily price (in USD) of AAVE |
| **TVL** | Value (in USD) of funds locked in the project's smart contracts |
| **Revenue** | The amount of revenue (in USD) that is distributed to AAVE holders |
| **Loan vol usd** | Daily volume (in USD) of Aave loans |
| **Deposit vol usd** | Daily volume (in USD) of Aave deposits |
| **Outstanding loan** | The value (in USD) of outstanding loans in Aave |
| **Outstanding deposit** | The value (in USD) of outstanding deposits in Aave |
| **Liquidation usd** | The value (in USD) of collateral liquidated daily in Aave |
| **AAVE holder** | The number of Ethereum addresses that have a non-zero balance of AAVE token |
| **Active user** | Daily active users of Aave protocol |
| **Developer** | Daily active developers of Aave protocol |

Note: This table introduces a series of factors related to Aave protocol. The factors can be retrieved from *Dune.xyz and tokenterminal.com*.

**Table 4. The effects of liquidity risks on Aave protocol**

| **Panel A: Liquidity** | | | | | | |
|---|---|---|---|---|---|---|
|  | (1) ΔMktC_F | (2) ΔMktC_C | **(3) Revenue** | (4) ΔTVL | (5) ΔAAVE | **(6) ΔAAVE holder** |
| **Liquidity** | -0.01 | -0.01 | **0.14*** | -0.01 | -0.01 | **-0.13*** |
|  | (-0.98) | (-0.97) | **(7.43)** | (-0.70) | (-0.98) | **(-11.09)** |
| **ΔOutstanding loan** | 0.07 | 0.08 | **-0.43*** | **0.17** | 0.07 | 0.01 |
|  | (1.04) | (1.15) | **(-0.43)** | **(2.31)** | (1.04) | (0.12) |
| **ΔOutstanding deposit** | 0.01 | 0.01 | **-0.28** | -0.02 | 0.01 | -0.03 |
|  | (0.20) | (0.11) | **(-2.11)** | (-0.27) | (0.20) | (-0.41) |
| **Deposits vol usd** | -0.04 | -0.04 | **0.41*** | -0.05 | -0.04 | **-0.12*** |
|  | (-1.23) | (-1.18) | **(6.46)** | (-1.62) | (-1.23) | **(-3.18)** |
| **Loan vol usd** | 0.01 | 0.00 | **0.17** | **0.06*** | 0.01 | **0.08*** |
|  | (0.27) | (0.10) | **(2.32)** | **(1.65)** | (0.27) | **(1.82)** |
| **Liquidation usd** | -0.03 | -0.06 | 0.01 | **-0.14*** | -0.03 | 0.04 |
|  | (-0.51) | (-1.14) | (0.11) | **(-2.45)** | (-0.51) | (0.61) |
| **ΔActive user** | 0.03 | 0.04 | 0..01 | **0.05*** | 0.03 | 0.03 |
|  | (1.41) | (1.51) | (0.12) | **(1.92)** | (1.41) | (0.95) |
| ΔDeveloper | -0.03 | -0.03 | -0.01 | -0.01 | -0.03 | 0.00 |
|  | (-1.44) | (-1.26) | (-0.16) | (-0.61) | (-1.44) | (0.17) |
| N | 792 | 792 | 792 | 789 | 792 | 792 |
| Adj R-sq | 0.00 | 0.01 | 0.18 | 0.03 | 0.00 | 0.14 |
| **Panel B: Utilization** | | | | | | |
|  | (1) ΔMktC_F | (2) ΔMktC_C | **(3) Revenue** | (4) ΔTVL | (5) ΔAAVE | (6) ΔAAVE holder |
| **Utilization** | 0.00 | 0.00 | **0.37*** | 0.02 | 0.00 | 0.02 |
|  | (-0.26) | (-0.17) | **(13.47)** | (1.18) | (-0.26) | (1.18) |
| **ΔOutstanding loan** | 0.07 | 0.08 | **-0.59*** | **0.16*** | 0.07 | 0.02 |
|  | (1.08) | (1.18) | **(-4.27)** | **(2.23)** | (1.08) | (0.23) |
| ΔOutstanding deposit | 0.01 | 0.01 | -0.17 | -0.01 | 0.01 | -0.03 |
|  | (0.18) | (0.10) | (-1.40) | (-0.20) | (0.18) | (-0.37) |
| **Deposits vol usd** | -0.03 | -0.03 | **0.26*** | **-0.06*** | -0.03 | **-0.12*** |
|  | (-1.15) | (-1.12) | **(4.30)** | **(-1.79)** | (-1.15) | **(-2.99)** |
| **Loan vol usd** | 0.01 | 0.00 | **0.23*** | **0.06*** | 0.01 | 0.04 |
|  | (0.19) | (0.02) | **(3.54)** | **(1.63)** | (0.19) | (0.88) |
| **Liquidation usd** | -0.02 | -0.06 | -0.05 | **-0.14*** | -0.02 | 0.06 |
|  | (-0.47) | (-1.10) | (-0.51) | **(-2.47)** | (-0.47) | (0.88) |
| **ΔActive user** | 0.03 | 0.04 | 0.01 | **0.05*** | 0.03 | 0.02 |
|  | (1.39) | (1.50) | (0.26) | **(1.91)** | (1.40) | (0.72) |
| ΔDeveloper | -0.03 | -0.03 | -0.02 | -0.01 | -0.03 | 0.01 |
|  | (-1.41) | (-1.24) | (-0.43) | (-0.60) | (-1.41) | (0.43) |
| N | 792 | 792 | 792 | 789 | 792 | 792 |
| Adj R-sq | 0.00 | 0.00 | 0.29 | 0.03 | 0.00 | 0.00 |
| **Panel C: Repeat deposit ratio** | | | | | | |
|  | (1) ΔMktC_F | (2) ΔMktC_C | (3) Revenue | (4) ΔTVL | (5) ΔAAVE | (6) ΔAAVE holder |
| Repeat deposit ratio | 0.01 | 0.00 | 0.05 | 0.02 | 0.01 | 0.01 |
|  | (0.33) | (0.20) | (1.25) | (0.87) | (0.32) | (0.37) |
| **ΔOutstanding loan** | 0.07 | 0.08 | **-0.45*** | **0.17*** | 0.07 | 0.03 |
|  | (1.06) | (1.17) | **(-2.95)** | **(2.32)** | (1.06) | (0.32) |
| ΔOutstanding deposit | 0.01 | 0.01 | **-0.27\*\*** | -0.02 | 0.01 | -0.04 |
| | (0.20) | (0.11) | **(-1.99)** | (-0.26) | (0.20) | (-0.43) |
| Deposits vol usd | -0.04 | -0.04 | **0.41\*\*\*** | -0.05 | -0.04 | **-0.12\*\*\*** |
| | (-1.22) | (-1.17) | **(6.16)** | (-1.60) | (-1.22) | **(-2.82)** |
| Loan vol usd | 0.01 | 0.00 | **0.22\*\*\*** | **0.06\*** | 0.01 | 0.04 |
| | (0.22) | (0.04) | **(2.96)** | **(1.68)** | (0.22) | (0.88) |
| Liquidation usd | -0.02 | -0.06 | -0.01 | **-0.13\*\*** | -0.02 | 0.07 |
| | (-0.47) | (-1.10) | (-0.09) | **(-2.40)** | (-0.47) | (0.92) |
| ΔActive user | 0.03 | 0.04 | 0.01 | **0.05\*** | 0.03 | 0.02 |
| | (1.38) | (1.49) | (0.17) | **(1.87)** | (1.38) | (0.70) |
| ΔDeveloper | -0.03 | -0.03 | -0.02 | -0.02 | -0.03 | 0.01 |
| | (-1.44) | (-1.25) | (-0.45) | (-0.67) | (-1.44) | (0.40) |
| N | 792 | 792 | 792 | 789 | 792 | 792 |
| Adj R-sq | 0.00 | 0.00 | 0.13 | 0.03 | 0.00 | 0.00 |
| **Panel D: Repeat loan ratio** | | | | | | |
| | **(1)** | **(2)** | **(3)** | (4) | **(5)** | **(6)** |
| | **ΔMktC_F** | **ΔMktC_C** | **Revenue** | ΔTVL | **ΔAAVE** | **ΔAAVE holder** |
| **Repeat loan ratio** | **0.03\*** | **0.03\*** | **-0.07\*\*** | 0.01 | **0.03\*** | **0.04\*** |
| | **(1.93)** | **(1.70)** | **(-1.96)** | (0.73) | **(1.93)** | **(1.68)** |
| **ΔOutstanding loan** | 0.07 | 0.08 | **-0.45\*\*\*** | **0.17\*\*** | 0.07 | 0.03 |
| | (1.06) | (1.17) | **(-2.95)** | **(2.32)** | (1.06) | (0.32) |
| **ΔOutstanding deposit** | 0.01 | 0.00 | **-0.26\*\*** | -0.02 | 0.01 | -0.04 |
| | (0.13) | (0.04) | **(-1.94)** | (-0.30) | (0.13) | (-0.50) |
| **Deposits vol usd** | -0.03 | -0.03 | **0.39\*\*\*** | -0.05 | -0.03 | **-0.11\*\*\*** |
| | (-1.06) | (-1.02) | **(5.98)** | (-1.54) | (-1.06) | **(-2.68)** |
| **Loan vol usd** | 0.01 | 0.01 | **0.20\*\*\*** | **0.06\*** | 0.01 | 0.05 |
| | (0.37) | (0.18) | **(2.66)** | **(1.66)** | (0.37) | (1.00) |
| **Liquidation usd** | -0.02 | -0.05 | -0.03 | **-0.13\*\*** | -0.02 | 0.08 |
| | (-0.32) | (-0.96) | (-0.29) | **(-2.36)** | (-0.32) | (1.05) |
| **ΔActive user** | 0.03 | 0.04 | 0.01 | **0.05\*** | 0.03 | 0.02 |
| | (1.41) | (1.51) | (0.22) | **(1.92)** | (1.41) | (0.72) |
| **ΔDeveloper** | -0.03 | -0.03 | -0.01 | -0.01 | -0.03 | 0.01 |
| | (-1.58) | (-1.39) | (-0.17) | (-0.66) | (-1.58) | (0.28) |
| N | 792 | 792 | 792 | 789 | 792 | 792 |
| Adj R-sq | 0.01 | 0.01 | 0.13 | 0.03 | 0.01 | 0.01 |

Note: This table reports regression results for influence of liquidity risks on Aave protocol. In columns (1) – (6) of each panel, the dependent variable is $\Delta MktC\_F, \Delta MktC\_C, Revenue, \Delta TVL, \Delta AAVE$, and $\Delta AAVE\ holder$ respectively. T-statistics are reported in parentheses. *, **, and *** denote significance levels at the 10%, 5%, and 1% levels based on the standard t-statistics.

**Table 5. The effects of liquidity risks on Compound protocol**

| **Panel A: Liquidity** | | | | | | |
|---|---|---|---|---|---|---|
| | **(1)** | **(2)** | (3) | (4) | **(5)** | **(6)** |
| | **ΔMktC_F** | **ΔMktC_C** | ΔRevenue | ΔTVL | **ΔCOMP** | **ΔCOMP holder** |
| **Liquidity** | **-0.02\*** | **-0.02\*** | 0.00 | -0.01 | **-0.02\*** | **-0.03\*\*\*** |
| | **(-1.69)** | **(-1.73)** | (0.66) | (-1.13) | **(-1.69)** | **(-3.90)** |
| **ΔOutstanding loan** | **-0.08\*** | -0.06 | 0.01 | -0.05 | **-0.07\*** | -0.01 |
| | **(-1.80)** | (-1.56) | (0.29) | (-1.42) | **(-1.80)** | (-0.42) |
| **ΔOutstanding deposit** | **0.43\*\*\*** | **0.42\*\*\*** | **0.23\*\*\*** | **0.35\*\*\*** | **0.43\*\*\*** | 0.05 |
| | **(7.45)** | **(7.75)** | **(6.47)** | **(7.25)** | **(7.45)** | (1.30) |
| **Loan vol usd** | **-0.11\*\*** | **-0.10\*\*** | -0.05 | 0.02 | **-0.11\*\*** | **0.10\*\*\*** |
| | **(-2.22)** | **(-2.07)** | (-1.61) | (0.45) | **(-2.22)** | **(3.17)** |
| Liquidation usd | 0.07 | 0.11 | -0.02 | 0.00 | 0.07 | 0.01 |
| | (0.98) | (1.55) | (-0.58) | (0.01) | (0.98) | (0.28) |
| **Active user** | -0.01 | -0.01 | 0.01 | 0.01 | -0.01 | **0.31\*\*\*** |
| | (-0.41) | (-0.44) | (0.42) | (0.42) | (-0.41) | **(18.78)** |
| **ΔDeveloper** | -0.03 | -0.02 | **-0.05\*\*\*** | -0.02 | -0.03 | 0.02 |
| | (-0.86) | (-0.65) | **(-3.00)** | (-0.74) | (-0.86) | (0.97) |
| N | 558 | 558 | 641 | 641 | 558 | 624 |
| Adj R-sq | 0.10 | 0.11 | 0.08 | 0.08 | 0.10 | 0.45 |
| **Panel B: Utilization** | | | | | | |
| | (1) | (2) | (3) | (4) | (5) | (6) |
| | ΔMktC_F | ΔMktC_C | ΔRevenue | ΔTVL | ΔCOMP | ΔCOMP holder |
| Utilization | 0.00 | 0.01 | 0.00 | -0.01 | 0.00 | 0.01 |
| | (0.21) | (0.35) | (0.37) | (-0.73) | (0.21) | (0.87) |
| **ΔOutstanding loan** | **-0.08\*** | -0.06 | 0.01 | -0.05 | **-0.08\*** | -0.01 |
| | **(-1.83)** | (-1.59) | (0.30) | (-1.44) | **(-1.83)** | (-0.48) |
| **ΔOutstanding deposit** | **0.44\*\*\*** | **0.43\*\*\*** | **0.23\*\*\*** | **0.35\*\*\*** | **0.44\*\*\*** | **0.07\*** |
| | **(7.59)** | **(7.90)** | **(6.43)** | **(7.27)** | **(7.59)** | **(1.79)** |
| **Loan vol usd** | **-0.10\*\*** | **-0.09\*** | **-0.05\*** | 0.03 | **-0.10\*\*** | **0.10\*\*\*** |

|  | (-1.97) | (-1.85) | (-1.66) | (0.69) | (-1.97) | (2.84) |
|---|---|---|---|---|---|---|
| **Liquidation usd** | 0.09 | **0.12*** | -0.03 | 0.01 | 0.09 | 0.03 |
|  | (1.20) | **(1.78)** | (-0.63) | (0.10) | (1.20) | (0.75) |
| **Active user** | 0.01 | 0.01 | 0.00 | 0.02 | 0.01 | **0.33*** |
|  | (0.39) | (0.40) | (0.25) | (0.77) | (0.39) | **(21.23)** |
| **ΔDeveloper** | -0.02 | -0.02 | **-0.05*** | -0.02 | -0.02 | 0.02 |
|  | (-0.77) | (-0.55) | **(-3.03)** | (-0.71) | (-0.77) | (1.29) |
| N | 558 | 558 | 641 | 641 | 558 | 624 |
| Adj R-sq | 0.09 | 0.10 | 0.08 | 0.08 | 0.09 | 0.43 |

**Panel C: Repeat deposit ratio**

|  | (1) | (2) | (3) | (4) | (5) | (6) |
|---|---|---|---|---|---|---|
|  | ΔMktC_F | ΔMktC_C | ΔRevenue | ΔTVL | ΔCOMP | ΔCOMP holder |
| Repeat deposit ratio | 0.00 | 0.00 | 0.00 | -0.01 | 0.00 | -0.02 |
|  | (0.15) | (0.14) | (-0.13) | (-0.36) | (0.15) | (-1.51) |
| **ΔOutstanding loan** | **-0.08*** | -0.06 | 0.01 | -0.05 | **-0.08*** | -0.01 |
|  | **(-1.83)** | (-1.60) | (0.30) | (-1.43) | **(-1.83)** | (-0.45) |
| **ΔOutstanding deposit** | **0.44*** | **0.43*** | **0.23*** | **0.36*** | **0.44*** | **0.06*** |
|  | **(7.61)** | **(7.91)** | **(6.44)** | **(7.41)** | **(7.61)** | **(1.71)** |
| **Loan vol usd** | **-0.10*** | **-0.09*** | **-0.05*** | 0.02 | **-0.10*** | **0.10*** |
|  | **(-1.98)** | **(-1.82)** | **(-1.63)** | (0.47) | **(-1.98)** | **(3.26)** |
| **Liquidation usd** | 0.09 | **0.12*** | -0.03 | 0.01 | 0.09 | 0.03 |
|  | (1.19) | **(1.77)** | (-0.65) | (0.13) | (1.19) | (0.71) |
| **Active user** | 0.01 | 0.01 | 0.00 | 0.02 | 0.01 | **0.33*** |
|  | (0.37) | (0.35) | (0.21) | (0.85) | (0.37) | **(21.25)** |
| **ΔDeveloper** | -0.02 | -0.02 | **-0.05*** | -0.02 | -0.02 | 0.02 |
|  | (-0.77) | (-0.56) | **(-3.04)** | (-0.67) | (-0.77) | (1.26) |
| N | 558 | 558 | 641 | 641 | 558 | 624 |
| Adj R-sq | 0.09 | 0.10 | 0.08 | 0.08 | 0.09 | 0.43 |

**Panel D: Repeat loan ratio**

|  | (1) | (2) | (3) | (4) | (5) | (6) |
|---|---|---|---|---|---|---|
|  | ΔMktC_F | ΔMktC_C | ΔRevenue | ΔTVL | ΔCOMP | ΔCOMP holder |
| Repeat loan ratio | 0.03 | 0.02 | 0.00 | 0.01 | 0.03 | 0.02 |
|  | (1.24) | (0.98) | (0.07) | (0.49) | (1.24) | (1.25) |
| **ΔOutstanding loan** | **-0.08*** | -0.06 | 0.01 | -0.05 | **-0.08*** | -0.01 |
|  | **(-1.84)** | (-1.60) | (0.30) | (-1.44) | **(-1.84)** | (-0.48) |
| **ΔOutstanding deposit** | **0.43*** | **0.42*** | **0.23*** | **0.35*** | **0.43*** | 0.06 |
|  | **(7.42)** | **(7.75)** | **(6.41)** | **(7.34)** | **(7.42)** | (1.58) |
| **Loan vol usd** | **-0.10*** | **-0.08*** | **-0.05*** | 0.02 | **-0.10*** | **0.11*** |
|  | **(-1.87)** | **(-1.73)** | **(-1.63)** | (0.50) | **(-1.87)** | **(3.37)** |
| **Liquidation usd** | 0.09 | **0.12*** | -0.03 | 0.01 | 0.09 | 0.03 |
|  | (1.24) | **(1.81)** | (-0.65) | (0.15) | (1.24) | (0.76) |
| **Active user** | 0.01 | 0.01 | 0.00 | 0.02 | 0.01 | **0.33*** |
|  | (0.39) | (0.37) | (0.21) | (0.85) | (0.39) | **(21.26)** |
| **ΔDeveloper** | -0.02 | -0.02 | **-0.05*** | -0.02 | -0.02 | 0.02 |
|  | (-0.77) | (-0.56) | **(-3.04)** | (-0.68) | (-0.77) | (1.26) |
| N | 558 | 558 | 641 | 641 | 558 | 624 |
| Adj R-sq | 0.10 | 0.10 | 0.08 | 0.08 | 0.10 | 0.43 |

Note: This table reports regression results for influence of liquidity risks on Compound protocol. In columns (1) – (6) of each panel, the dependent variable is $\Delta MktC\_F, \Delta MktC\_C, \Delta Revenue, \Delta TVL, \Delta COMP$, and $\Delta COMP\ holder$ respectively. T-statistics are reported in parentheses. *, **, and *** denote significance levels at the 10%, 5%, and 1% levels based on the standard t-statistics.

**Table 6. The effects of liquidity risks and Aave v2 on Aave protocol**

**Panel A: Liquidity**

|  | (1) | (2) | **(3)** | (4) | (5) | **(6)** |
|---|---|---|---|---|---|---|
|  | ΔMktC_F | ΔMktC_C | **Revenue** | ΔTVL | ΔAAVE | **ΔAAVE holder** |
| **Liquidity** | -0.01 | -0.01 | **0.14*** | -0.01 | -0.01 | **-0.13*** |
|  | (-0.96) | (-0.95) | **(7.33)** | (-0.70) | (-0.96) | **(-11.03)** |
| Aave v2 | -0.01 | -0.01 | 0.08 | 4.62e+09 | -0.01 | -0.09 |
|  | (-0.15) | (-0.14) | (0.85) | (0.01) | (-0.15) | (-0.17) |
| **ΔOutstanding loan** | 0.07 | 0.08 | **-0.43*** | **0.17*** | 0.07 | 0.01 |
|  | (1.04) | (1.15) | **(-2.89)** | **(2.31)** | (1.04) | (0.12) |
| **ΔOutstanding deposit** | 0.01 | 0.01 | **-0.28*** | -0.02 | 0.01 | -0.03 |
|  | (0.20) | (0.11) | **(-2.12)** | (-0.27) | (0.20) | (-0.41) |
| **Deposits vol usd** | -0.04 | -0.04 | **0.41*** | -0.05 | -0.04 | **-0.12*** |
|  | (-1.23) | (-1.18) | **(6.44)** | (-1.62) | (-1.23) | **(-3.17)** |
| **Loan vol usd** | 0.01 | 0.00 | **0.17*** | **0.06*** | 0.01 | **0.08*** |
|  | (0.28) | (0.10) | **(2.31)** | **(1.65)** | (0.28) | **(1.82)** |
| **Liquidation usd** | -0.03 | -0.06 | 0.01 | **-0.14*** | -0.03 | 0.04 |
|  | (-0.51) | (-1.14) | (0.11) | **(-2.45)** | (-0.51) | (0.61) |

|  | (1) ΔMktC_F | (2) ΔMktC_C | (3) Revenue | (4) ΔTVL | (5) ΔAAVE | (6) ΔAAVE holder |
|---|---|---|---|---|---|---|
| ΔActive user | 0.03 | 0.04 | 0.01 | **0.05*** | 0.03 | 0.03 |
|  | (1.41) | (1.51) | (0.12) | **(1.92)** | (1.41) | (0.94) |
| ΔDeveloper | -0.03 | -0.03 | -0.01 | -0.01 | -0.03 | 0.00 |
|  | (-1.44) | (-1.27) | (-0.13) | (-0.61) | (-1.44) | (0.17) |
| N | 792 | 792 | 792 | 789 | 792 | 792 |
| Adj R-sq | 0.00 | 0.00 | 0.18 | 0.03 | 0.00 | 0.14 |

**Panel B: Utilization**

|  | (1) ΔMktC_F | (2) ΔMktC_C | **(3) Revenue** | (4) ΔTVL | (5) ΔAAVE | (6) ΔAAVE holder |
|---|---|---|---|---|---|---|
| **Utilization** | 0.00 | 0.00 | **0.36*** | 0.02 | 0.00 | 0.02 |
|  | (-0.26) | (-0.16) | **(13.42)** | (1.17) | (-0.26) | (1.22) |
| Aave v2 | -0.01 | -0.01 | 0.09 | -5.18e+10 | -0.01 | -0.06 |
|  | (-0.23) | (-0.22) | (1.07) | (-0.07) | (-0.23) | (-1.06) |
| **ΔOutstanding loan** | 0.07 | 0.08 | **-0.59*** | **0.16**** | 0.07 | 0.02 |
|  | (1.08) | (1.17) | **(-4.26)** | **(2.20)** | (1.08) | (0.22) |
| ΔOutstanding deposit | 0.01 | 0.01 | -0.17 | -0.01 | 0.01 | -0.03 |
|  | (0.18) | (0.10) | (-1.40) | (-0.20) | (0.18) | (-0.37) |
| **Deposits vol usd** | -0.03 | -0.03 | **0.26*** | **-0.06*** | -0.03 | **-0.12**** |
|  | (-1.15) | (-1.12) | **(4.29)** | **(-1.78)** | (-1.15) | **(-2.98)** |
| **Loan vol usd** | 0.01 | 0.00 | **0.23*** | 0.06 | 0.01 | 0.04 |
|  | (0.19) | (0.02) | **(3.51)** | (1.62) | (0.19) | (0.90) |
| **Liquidation usd** | -0.02 | -0.06 | -0.05 | **-0.14**** | -0.02 | 0.06 |
|  | (-0.47) | (-1.10) | (-0.51) | **(-2.44)** | (-0.47) | (0.87) |
| **ΔActive user** | 0.03 | 0.04 | 0.01 | **0.05*** | 0.03 | 0.02 |
|  | (1.39) | (1.50) | (0.26) | **(1.90)** | (1.39) | (0.72) |
| ΔDeveloper | -0.03 | -0.03 | -0.02 | -0.01 | -0.03 | 0.01 |
|  | (-1.42) | (-1.24) | (-0.40) | (-0.60) | (-1.42) | (0.39) |
| N | 792 | 792 | 792 | 789 | 792 | 792 |
| Adj R-sq | 0.00 | 0.00 | 0.29 | 0.01 | 0.00 | 0.01 |

**Panel C: Repeat deposit ratio**

|  | (1) ΔMktC_F | (2) ΔMktC_C | (3) Revenue | (4) ΔTVL | (5) ΔAAVE | (6) ΔAAVE holder |
|---|---|---|---|---|---|---|
| Repeat deposit ratio | 0.01 | 0.00 | 0.04 | 0.02 | 0.01 | 0.01 |
|  | (0.34) | (0.21) | (1.17) | (0.89) | (0.34) | (0.43) |
| Aave v2 | -0.01 | -0.01 | 0.13 | -1.42e+11 | -0.01 | -0.06 |
|  | (-0.25) | (-0.23) | (1.34) | (-0.19) | (-0.25) | (-1.04) |
| **ΔOutstanding loan** | 0.07 | 0.08 | **-0.45*** | **0.17**** | 0.07 | 0.03 |
|  | (1.06) | (1.17) | **(-2.94)** | **(2.32)** | (1.06) | (0.31) |
| **ΔOutstanding deposit** | 0.01 | 0.01 | **-0.27**** | -0.02 | 0.01 | -0.04 |
|  | (0.20) | (0.11) | **(-1.99)** | (-0.26) | (0.20) | (-0.43) |
| **Deposits vol usd** | -0.04 | -0.04 | **0.40*** | -0.05 | -0.04 | **-0.11**** |
|  | (-1.21) | (-1.16) | **(6.14)** | (-1.60) | (-1.21) | **(-2.81)** |
| **Loan vol usd** | 0.01 | 0.00 | **0.22*** | **0.06*** | 0.01 | 0.04 |
|  | (0.23) | (0.04) | **(2.92)** | **(1.68)** | (0.23) | (0.90) |
| **Liquidation usd** | -0.02 | -0.06 | -0.01 | **-0.13**** | -0.02 | 0.07 |
|  | (-0.47) | (-1.10) | (-0.09) | **(-2.40)** | (-0.47) | (0.92) |
| **ΔActive user** | 0.03 | 0.04 | 0.01 | **0.05*** | 0.03 | 0.02 |
|  | (1.38) | (1.49) | (0.17) | **(1.87)** | (1.38) | (0.69) |
| ΔDeveloper | -0.03 | -0.03 | -0.02 | -0.02 | -0.03 | 0.01 |
|  | (-1.44) | (-1.26) | (-0.40) | (-0.67) | (-1.44) | (0.36) |
| N | 792 | 792 | 792 | 789 | 792 | 792 |
| Adj R-sq | 0.00 | 0.00 | 0.13 | 0.03 | 0.00 | 0.00 |

**Panel D: Repeat loan ratio**

|  | **(1) ΔMktC_F** | **(2) ΔMktC_C** | **(3) Revenue** | (4) ΔTVL | **(5) ΔAAVE** | **(6) ΔAAVE holder** |
|---|---|---|---|---|---|---|
| **Repeat loan ratio** | **0.03**** | **0.03*** | **-0.07**** | 0.01 | **0.03**** | **0.04*** |
|  | **(1.93)** | **(1.71)** | **(-2.00)** | (0.72) | **(1.93)** | **(1.71)** |
| Aave v2 | -0.01 | -0.01 | 0.14 | -1.72e+10 | -0.01 | -0.06 |
|  | (-0.28) | (-0.26) | (1.46) | (-0.02) | (-0.28) | (-1.05) |
| **ΔOutstanding loan** | 0.07 | 0.08 | **-0.45*** | **0.17**** | 0.07 | 0.03 |
|  | (1.06) | (1.16) | **(-2.94)** | **(2.29)** | (1.06) | (0.31) |
| **ΔOutstanding deposit** | 0.01 | 0.00 | **-0.26**** | -0.02 | 0.01 | -0.04 |
|  | (0.13) | (0.04) | **(-1.94)** | (-0.30) | (0.13) | (-0.50) |
| **Deposits vol usd** | -0.03 | -0.03 | **0.39*** | -0.05 | -0.03 | **-0.11**** |
|  | (-1.05) | (-1.02) | **(5.96)** | (-1.52) | (-1.05) | **(-2.66)** |
| **Loan vol usd** | 0.01 | 0.01 | **0.19*** | **0.06*** | 0.01 | 0.05 |
|  | (0.38) | (0.18) | **(2.63)** | **(1.64)** | (0.38) | (1.02) |
| **Liquidation usd** | -0.02 | -0.05 | -0.03 | **-0.13**** | -0.02 | 0.08 |
|  | (-0.32) | (-0.96) | (-0.29) | **(-2.33)** | (-0.32) | (1.05) |
| **ΔActive user** | 0.03 | 0.04 | 0.01 | **0.05*** | 0.03 | 0.02 |
|  | (1.41) | (1.51) | (0.22) | **(1.89)** | (1.41) | (0.72) |
| ΔDeveloper | -0.03 | -0.03 | -0.01 | -0.01 | -0.03 | 0.01 |
|  | (-1.59) | (-1.39) | (-0.12) | (-0.65) | (-1.59) | (0.25) |
| N | 792 | 792 | 792 | 789 | 792 | 792 |

|  | Adj R-sq | 0.01 | 0.01 | 0.13 | 0.00 | 0.01 | 0.01 |

Note: This table reports regression results for influence of liquidity risks and Aave v2 on Aave protocol. In columns (1) – (6) of each panel, the dependent variable is $\Delta MktC\_F, \Delta MktC\_C, Revenue, \Delta TVL, \Delta AAVE$, and $\Delta AAVE\ holder$ respectively. T-statistics are reported in parentheses. *, **, and *** denote significance levels at the 10%, 5%, and 1% levels based on the standard t-statistics.

**Table 7. The effects of liquidity risks and Compound v3 on Compound protocol**

| **Panel A: Liquidity** | | | | | | |
|---|---|---|---|---|---|---|
|  | **(1)** ΔMktC_F | **(2)** ΔMktC_C | (3) ΔRevenue | (4) ΔTVL | **(5)** ΔCOMP | **(6)** ΔCOMP holder |
| **Liquidity** | **-0.02*** | **-0.02*** | 0.01 | -0.01 | **-0.02*** | **-0.03**** |
|  | **(-1.71)** | **(-1.76)** | (0.70) | (-1.12) | **(-1.71)** | **(-4.10)** |
| Compound v3 | 0.00 | 0.00 | 0.00 | 0.00 | 0.00 | 0.01 |
|  | (0.25) | (0.32) | (-0.24) | (0.11) | (0.25) | (1.31) |
| **ΔOutstanding loan** | **-0.07*** | -0.06 | 0.01 | -0.05 | **-0.07*** | -0.01 |
|  | **(-1.80)** | (-1.56) | (0.28) | (-1.42) | **(-1.80)** | (-0.40) |
| **ΔOutstanding deposit** | **0.43**** | **0.42**** | **0.24**** | **0.35**** | **0.43**** | 0.04 |
|  | **(7.40)** | **(7.69)** | **(6.46)** | **(7.21)** | **(7.40)** | (1.19) |
| **Loan vol usd** | **-0.11**** | **-0.10**** | -0.05* | 0.02 | **-0.11**** | **0.11**** |
|  | **(-2.17)** | **(-2.10)** | (-1.63) | (0.46) | **(-2.17)** | **(3.31)** |
| Liquidation usd | 0.07 | 0.11 | -0.02 | 0.00 | 0.07 | 0.01 |
|  | (0.98) | (1.56) | (-0.58) | (0.01) | (0.98) | (0.28) |
| Active user | -0.01 | -0.01 | 0.01 | 0.01 | -0.01 | **0.31**** |
|  | (-0.40) | (-0.42) | (0.41) | (0.43) | (-0.40) | **(18.84)** |
| **ΔDeveloper** | -0.03 | -0.02 | **-0.05**** | -0.02 | -0.02 | 0.02 |
|  | (-0.86) | (-0.64) | **(-3.00)** | (-0.74) | (-0.86) | (0.98) |
| N | 558 | 558 | 641 | 641 | 558 | 624 |
| Adj R-sq | 0.10 | 0.11 | 0.08 | 0.08 | 0.10 | 0.45 |
| **Panel B: Utilization** | | | | | | |
|  | (1) ΔMktC_F | (2) ΔMktC_C | (3) ΔRevenue | (4) ΔTVL | (5) ΔCOMP | (6) ΔCOMP holder |
| Utilization | 0.00 | 0.01 | 0.00 | -0.01 | 0.00 | 0.01 |
|  | (0.19) | (0.35) | (0.36) | (-0.78) | (0.19) | (0.94) |
| Compound v3 | 0.00 | 0.00 | 0.00 | 0.00 | 0.00 | 0.00 |
|  | (-0.06) | (0.04) | (0.01) | (-0.34) | (-0.06) | (0.44) |
| **ΔOutstanding loan** | **-0.08*** | -0.06 | 0.01 | -0.05 | **-0.08*** | -0.01 |
|  | **(-1.83)** | (-1.59) | (0.30) | (-1.44) | **(-1.83)** | (-0.47) |
| **ΔOutstanding deposit** | **0.44**** | **0.43**** | **0.23**** | **0.35**** | **0.44**** | **0.07*** |
|  | **(7.58)** | **(7.89)** | **(6.42)** | **(7.27)** | **(7.58)** | **(1.78)** |
| **Loan vol usd** | **-0.10**** | **-0.09*** | -0.05* | 0.03 | **-0.10**** | **0.10**** |
|  | **(-1.96)** | **(-1.84)** | (-1.66) | (0.67) | **(-1.96)** | **(2.86)** |
| **Liquidation usd** | 0.09 | **0.12*** | -0.03 | 0.00 | 0.09 | 0.03 |
|  | (1.19) | **(1.78)** | (-0.63) | (0.08) | (1.19) | (0.77) |
| Active user | 0.01 | 0.01 | 0.00 | 0.01 | 0.01 | **0.33**** |
|  | (0.37) | (0.40) | (0.24) | (0.70) | (0.37) | **(20.97)** |
| **ΔDeveloper** | -0.02 | -0.02 | **-0.05**** | -0.02 | -0.02 | 0.02 |
|  | (-0.77) | (-0.55) | **(-3.02)** | (-0.72) | (-0.77) | (1.30) |
| N | 558 | 558 | 641 | 641 | 558 | 624 |
| Adj R-sq | 0.09 | 0.10 | 0.08 | 0.08 | 0.09 | 0.43 |
| **Panel C: Repeat deposit ratio** | | | | | | |
|  | (1) ΔMktC_F | (2) ΔMktC_C | (3) ΔRevenue | (4) ΔTVL | (5) ΔCOMP | (6) ΔCOMP holder |
| Repeat deposit ratio | 0.00 | 0.00 | 0.00 | -0.01 | 0.00 | -0.02 |
|  | (0.16) | (0.15) | (-0.12) | (-0.34) | (0.16) | (-1.56) |
| Compound v3 | 0.00 | 0.00 | 0.00 | 0.00 | 0.00 | 0.00 |
|  | (-0.13) | (-0.07) | (-0.04) | (-0.14) | (-0.13) | (0.47) |
| **ΔOutstanding loan** | **-0.08*** | -0.06 | 0.01 | -0.05 | **-0.08*** | -0.01 |
|  | **(-1.83)** | (-1.60) | (0.30) | (-1.43) | **(-1.83)** | (-0.44) |
| **ΔOutstanding deposit** | **0.44**** | **0.43**** | **0.23**** | **0.36**** | **0.44**** | **0.06*** |
|  | **(7.59)** | **(7.89)** | **(6.42)** | **(7.40)** | **(7.59)** | **(1.69)** |
| **Loan vol usd** | **-0.10**** | **-0.09*** | -0.05 | 0.02 | **-0.10**** | **0.11**** |
|  | **(-1.98)** | **(-1.81)** | (-1.62) | (0.45) | **(-1.97)** | **(3.29)** |
| **Liquidation usd** | 0.09 | **0.12*** | -0.03 | 0.01 | 0.09 | 0.03 |
|  | (1.18) | **(1.76)** | (-0.65) | (0.13) | (1.18) | (0.72) |
| **Active user** | 0.01 | 0.01 | 0.00 | 0.02 | 0.01 | **0.33**** |
|  | (0.34) | (0.34) | (0.20) | (0.82) | (0.34) | **(21.07)** |
| **ΔDeveloper** | -0.02 | -0.02 | **-0.05**** | -0.02 | -0.02 | 0.02 |
|  | (-0.77) | (-0.56) | **(-3.04)** | (-0.68) | (-0.77) | (1.28) |
| N | 558 | 558 | 641 | 641 | 558 | 624 |
| Adj R-sq | 0.09 | 0.10 | 0.08 | 0.08 | 0.09 | 0.43 |

**Panel D: Repeat loan ratio**

|  | (1) ΔMktC_F | (2) ΔMktC_C | (3) ΔRevenue | (4) ΔTVL | (5) ΔCOMP | (6) ΔCOMP holder |
|---|---|---|---|---|---|---|
| Repeat loan ratio | 0.03 | 0.02 | 0.00 | 0.01 | 0.03 | 0.01 |
|  | (1.26) | (0.99) | (0.07) | (0.52) | (1.26) | (1.23) |
| Compound v3 | 0.00 | 0.00 | 0.00 | 0.00 | 0.00 | 0.00 |
|  | (-0.28) | (-0.18) | (-0.07) | (-0.26) | (-0.28) | (0.09) |
| **ΔOutstanding loan** | **-0.08*** | -0.06 | 0.01 | -0.05 | **-0.08*** | -0.01 |
|  | **(-1.84)** | (-1.60) | (0.30) | (-1.44) | **(-1.84)** | (-0.48) |
| **ΔOutstanding deposit** | **0.43*** | **0.42*** | **0.23*** | **0.35*** | **0.43*** | 0.06 |
|  | **(7.42)** | **(7.74)** | **(6.40)** | **(7.34)** | **(7.42)** | (1.57) |
| **Loan vol usd** | **-0.10*** | **-0.08*** | -0.05 | 0.02 | **-0.10*** | **0.11**** |
|  | **(-1.89)** | **(-1.74)** | (-1.62) | (0.46) | **(-1.89)** | **(3.35)** |
| **Liquidation usd** | 0.09 | **0.12*** | -0.03 | 0.01 | 0.09 | 0.03 |
|  | (1.23) | **(1.80)** | (-0.65) | (0.14) | (1.23) | (0.76) |
| **Active user** | 0.01 | 0.01 | 0.00 | 0.02 | 0.01 | **0.33**** |
|  | (0.34) | (0.33) | (0.20) | (0.80) | (0.34) | **(21.01)** |
| **ΔDeveloper** | -0.02 | -0.02 | **-0.05**** | -0.02 | -0.02 | 0.02 |
|  | (-0.78) | (-0.56) | **(-3.04)** | (-0.69) | (-0.78) | (1.26) |
| N | 558 | 558 | 641 | 641 | 558 | 624 |
| Adj R-sq | 0.09 | 0.10 | 0.08 | 0.08 | 0.09 | 0.43 |

Note: This table reports regression results for influence of liquidity risks and Compound v3 on Compound protocol. In columns (1) – (6) of each panel, the dependent variable is $\Delta MktC\_F, \Delta MktC\_C, \Delta Revenue, \Delta TVL, \Delta COMP$, and $\Delta COMP\ holder$ respectively. T-statistics are reported in parentheses. *, **, and *** denote significance levels at the 10%, 5%, and 1% levels based on the standard t-statistics.

**Table 8. DeFi hacks**

| Date | Hacked protocol | Date | Hacked protocol | Date | Hacked protocol |
|---|---|---|---|---|---|
| 2020-02-18 | bZx | 2021-03-05 | Paid Network | 2022-01-27 | Qubit Finance |
| 2020-03-12 | DAOMaker | 2021-04-19 | EasyFi | 2022-02-02 | Wormhole Bridge |
| 2020-04-18 | Uniswap | 2021-04-28 | Uranium Finance | 2022-03-29 | Ronin Bridge |
| 2020-04-19 | dForce | 2021-05-19 | PancakeBunny | 2022-04-17 | Beanstalk Farms |
| 2020-09-14 | bZx | 2021-05-30 | Belt Finance | 2022-06-05 | Maiar Exchange |
| 2020-09-29 | Eminence | 2021-08-10 | Poly Network | 2022-06-24 | Horizon Bridge |
| 2020-10-26 | Harvest | 2021-09-30 | Compound | 2022-08-02 | Nomad Bridge |
| 2020-11-12 | Akropolis | 2021-10-27 | Cream Finance | 2022-09-21 | Wintermute |
| 2020-11-14 | Value DeFi | 2021-11-05 | bZx | 2022-10-06 | BNB Chain |
| 2020-11-21 | Pickle Finance | 2021-12-02 | BadgerDAO | 2022-10-11 | Mango Markets |

Note: This table presents influential DeFi hacks happened in 2020 – 2022.

**Table 9. The effects of liquidity risks and DeFi hacks on Aave protocol**

**Panel A: Liquidity**

|  | (1) ΔMktC_F | (2) ΔMktC_C | **(3) Revenue** | (4) ΔTVL | (5) ΔAAVE | **(6) ΔAAVE holder** |
|---|---|---|---|---|---|---|
| **Liquidity** | -0.01 | -0.01 | **0.14*** | -0.01 | -0.01 | **-0.13**** |
|  | (-0.98) | (-0.97) | **(7.50)** | (-0.70) | (-0.98) | **(-11.12)** |
| **Hack** | 0.00 | 0.00 | **0.06*** | 0.01 | 0.00 | **-0.02**** |
|  | (0.08) | (0.15) | **(5.13)** | (1.14) | (0.08) | **(-3.20)** |
| **ΔOutstanding loan** | 0.07 | 0.08 | **-0.41**** | **0.17*** | 0.07 | 0.00 |
|  | (1.05) | (1.15) | **(-2.80)** | **(2.34)** | (1.05) | (0.03) |
| **ΔOutstanding deposit** | 0.01 | 0.01 | **-0.28*** | -0.02 | 0.01 | -0.03 |
|  | (0.20) | (0.11) | **(-2.17)** | (-0.27) | (0.20) | (-0.40) |
| **Deposits vol usd** | -0.04 | -0.04 | **0.43*** | -0.05 | -0.04 | **-0.13**** |
|  | (-1.23) | (-1.17) | **(6.81)** | (-1.56) | (-1.23) | **(-3.36)** |
| **Loan vol usd** | 0.01 | 0.00 | **0.14*** | 0.06 | 0.01 | **0.09*** |
|  | (0.27) | (0.09) | **(1.93)** | (1.55) | (0.27) | **(2.09)** |
| **Liquidation usd** | -0.03 | -0.06 | -0.02 | **-0.14**** | -0.03 | 0.05 |
|  | (-0.52) | (-1.15) | (-0.16) | **(-2.51)** | (-0.52) | (0.78) |
| **ΔActive user** | 0.03 | 0.04 | 0.01 | **0.05*** | 0.03 | 0.03 |
|  | (1.41) | (1.51) | (0.18) | **(1.94)** | (1.41) | (0.91) |
| **ΔDeveloper** | -0.03 | -0.03 | -0.02 | -0.02 | -0.03 | 0.01 |
|  | (-1.44) | (-1.27) | (-0.54) | (-0.70) | (-1.44) | (0.41) |
| N | 792 | 792 | 792 | 789 | 792 | 792 |
| Adj R-sq | 0.00 | 0.00 | 0.21 | 0.03 | 0.00 | 0.15 |

**Panel B: Utilization**

|  | (1) ΔMktC_F | (2) ΔMktC_C | **(3) Revenue** | (4) ΔTVL | (5) ΔAAVE | (6) ΔAAVE holder |
|---|---|---|---|---|---|---|
| **Utilization** | 0.00 | 0.00 | **0.35*** | 0.02 | 0.00 | 0.03 |
|  | (-0.27) | (-0.18) | **(13.10)** | (1.07) | (-0.27) | (1.51) |
| **Hack** | 0.00 | 0.00 | **0.05*** | 0.01 | 0.00 | **-0.03*** |
|  | (0.10) | (0.16) | **(4.16)** | (1.02) | (0.10) | **(-3.22)** |
| **ΔOutstanding loan** | 0.07 | 0.08 | **-0.57*** | **0.17**** | 0.07 | 0.01 |
|  | (1.08) | (1.18) | **(-4.16)** | **(2.26)** | (1.08) | (0.12) |
| ΔOutstanding deposit | 0.01 | 0.01 | -0.18 | -0.01 | 0.01 | -0.03 |
|  | (0.18) | (0.09) | (-1.46) | (-0.21) | (0.18) | (-0.34) |
| **Deposits vol usd** | -0.03 | -0.03 | **0.28*** | **-0.06*** | -0.03 | **-0.13*** |
|  | (-1.14) | (-1.11) | **(4.63)** | **(-1.72)** | (-1.14) | **(-3.23)** |
| **Loan vol usd** | 0.01 | 0.00 | **0.21*** | 0.05 | 0.00 | 0.05 |
|  | (0.18) | (0.00) | **(3.20)** | (1.54) | (0.18) | (1.16) |
| **Liquidation usd** | -0.02 | -0.06 | -0.08 | **-0.14**** | -0.02 | 0.07 |
|  | (-0.48) | (-1.11) | (-0.73) | **(-2.51)** | (-0.48) | (1.04) |
| **ΔActive user** | 0.03 | 0.04 | 0.01 | **0.05*** | 0.03 | 0.02 |
|  | (1.40) | (1.50) | (0.31) | **(1.93)** | (1.400) | (0.69) |
| ΔDeveloper | -0.03 | -0.03 | -0.03 | -0.02 | -0.03 | 0.02 |
|  | (-1.42) | (-1.24) | (-0.74) | (-0.68) | (-1.41) | (0.67) |
| N | 792 | 792 | 792 | 789 | 792 | 792 |
| Adj R-sq | 0.00 | 0.00 | 0.30 | 0.03 | 0.00 | 0.02 |
| **Panel C: Repeat deposit ratio** | | | | | | |
|  | (1) ΔMktC_F | (2) ΔMktC_C | (3) Revenue | (4) ΔTVL | (5) ΔAAVE | (6) ΔAAVE holder |
| Repeat deposit ratio | 0.01 | 0.00 | 0.04 | 0.01 | 0.01 | 0.01 |
|  | (0.32) | (0.20) | (1.13) | (0.84) | (0.32) | (0.46) |
| **Hack** | 0.00 | 0.00 | **0.06*** | 0.01 | 0.00 | **-0.03*** |
|  | (0.06) | (0.14) | **(4.99)** | (1.12) | (0.06) | **(-3.09)** |
| **ΔOutstanding loan** | 0.07 | 0.08 | **-0.43*** | **0.17**** | 0.07 | 0.02 |
|  | (1.07) | (1.17) | **(-2.85)** | **(2.35)** | (1.07) | (0.24) |
| **ΔOutstanding deposit** | 0.01 | 0.01 | **-0.27**** | -0.02 | 0.01 | -0.04 |
|  | (0.20) | (0.11) | **(-2.04)** | (-0.26) | (0.20) | (-0.42) |
| **Deposits vol usd** | -0.04 | -0.04 | **0.42*** | -0.05 | -0.04 | **-0.12*** |
|  | (-1.21) | (-1.16) | **(6.51)** | (-1.54) | (-1.21) | **(-2.99)** |
| **Loan vol usd** | 0.01 | 0.00 | **0.19*** | 0.06 | 0.01 | 0.05 |
|  | (0.22) | (0.03) | **(2.56)** | (1.58) | (0.22) | (1.14) |
| **Liquidation usd** | -0.02 | -0.06 | -0.04 | **-0.14**** | -0.02 | 0.08 |
|  | (-0.47) | (-1.11) | (-0.36) | **(-2.46)** | (-0.47) | (1.09) |
| **ΔActive user** | 0.03 | 0.04 | 0.01 | **0.05*** | 0.03 | 0.02 |
|  | (1.38) | (1.49) | (0.23) | **(1.88)** | (1.38) | (0.66) |
| ΔDeveloper | -0.03 | -0.03 | -0.04 | -0.02 | -0.03 | 0.02 |
|  | (-1.44) | (-1.26) | (-0.81) | (-0.75) | (-1.44) | (0.62) |
| N | 792 | 792 | 792 | 789 | 792 | 792 |
| Adj R-sq | 0.00 | 0.00 | 0.15 | 0.03 | 0.00 | 0.01 |
| **Panel D: Repeat loan ratio** | | | | | | |
|  | **(1) ΔMktC_F** | **(2) ΔMktC_C** | **(3) Revenue** | (4) ΔTVL | **(5) ΔAAVE** | **(6) ΔAAVE holder** |
| **Repeat loan ratio** | **0.03*** | **0.03*** | **-0.07**** | 0.01 | **0.03*** | **0.04*** |
|  | **(1.92)** | **(1.70)** | **(-2.02)** | (0.72) | **(1.92)** | **(1.71)** |
| **Hack** | 0.00 | 0.00 | **0.07*** | 0.01 | 0.00 | **-0.03*** |
|  | (0.06) | (0.13) | **(5.05)** | (1.14) | (0.06) | **(-3.09)** |
| **ΔOutstanding loan** | 0.07 | 0.08 | **-0.43*** | **0.17**** | 0.07 | 0.02 |
|  | (1.06) | (1.17) | **(-2.85)** | **(2.35)** | (1.06) | (0.23) |
| **ΔOutstanding deposit** | 0.01 | 0.00 | **-0.26**** | -0.02 | 0.01 | -0.04 |
|  | (0.13) | (0.04) | **(-1.99)** | (-0.30) | (0.13) | (-0.49) |
| **Deposits vol usd** | -0.03 | -0.03 | **0.41*** | -0.05 | -0.03 | **-0.12*** |
|  | (-1.06) | (-1.02) | **(6.32)** | (-1.48) | (-1.06) | **(-2.85)** |
| **Loan vol usd** | 0.01 | 0.01 | **0.17**** | 0.06 | 0.01 | 0.06 |
|  | (0.36) | (0.16) | **(2.28)** | (1.56) | (0.36) | (1.26) |
| **Liquidation usd** | -0.02 | -0.05 | -0.07 | **-0.14**** | -0.02 | 0.09 |
|  | (-0.32) | (-0.97) | (-0.57) | **(-2.42)** | (-0.32) | (1.22) |
| **ΔActive user** | 0.03 | 0.04 | 0.01 | **0.05**** | 0.03 | 0.02 |
|  | (1.41) | (1.51) | (0.28) | **(1.93)** | (1.41) | (0.69) |
| ΔDeveloper | -0.03 | -0.03 | -0.03 | -0.02 | -0.03 | 0.01 |
|  | (-1.58) | (-1.39) | (-0.54) | (-0.74) | (-1.58) | (0.51) |
| N | 792 | 792 | 792 | 789 | 792 | 792 |
| Adj R-sq | 0.00 | 0.01 | 0.16 | 0.03 | 0.00 | 0.02 |

Note: This table reports regression results for influence of liquidity risks and DeFi hacks on Aave protocol. In columns (1) – (6) of each panel, the dependent variable is $\Delta MktC\_F, \Delta MktC\_C, Revenue, \Delta TVL, \Delta AAVE$, and $\Delta AAVE\ holder$ respectively. T-statistics are reported in parentheses. *, **, and *** denote significance levels at the 10%, 5%, and 1% levels based on the standard t-statistics.

**Table 10. The effects of liquidity risks and DeFi hacks on Compound protocol**

| Panel A: Liquidity | | | | | | |
|---|---|---|---|---|---|---|
| | (1) ΔMktC_F | (2) ΔMktC_C | (3) ΔRevenue | (4) ΔTVL | (5) ΔCOMP | (6) ΔCOMP holder |
| **Liquidity** | **-0.02*** | **-0.02*** | 0.00 | -0.01 | **-0.02*** | **-0.03**** |
| | **(-1.69)** | **(-1.73)** | (0.64) | (-1.12) | **(-1.69)** | **(-3.89)** |
| Hack | 0.00 | 0.00 | 0.00 | 0.00 | 0.00 | 0.00 |
| | (0.16) | (0.44) | (-0.59) | (0.33) | (0.16) | (0.67) |
| **ΔOutstanding loan** | **-0.07*** | -0.06 | 0.01 | -0.05 | **-0.07*** | -0.01 |
| | **(-1.79)** | (-1.54) | (0.26) | (-1.40) | **(-1.79)** | (-0.38) |
| **ΔOutstanding deposit** | **0.43**** | **0.42**** | **0.23**** | **0.35**** | **0.43**** | 0.05 |
| | **(7.45)** | **(7.76)** | **(6.45)** | **(7.25)** | **(7.45)** | (1.32) |
| **Loan vol usd** | **-0.11**** | **-0.10**** | -0.05 | 0.02 | **-0.11**** | **0.10**** |
| | **(-2.22)** | **(-2.07)** | (-1.62) | (0.45) | **(-2.22)** | **(3.16)** |
| Liquidation usd | 0.07 | 0.11 | -0.02 | 0.00 | 0.07 | 0.01 |
| | (0.99) | (1.57) | (-0.58) | (0.01) | (0.99) | (0.27) |
| **Active user** | -0.01 | -0.01 | 0.01 | 0.01 | -0.01 | **0.31**** |
| | (-0.43) | (-0.50) | (0.49) | (0.38) | (-0.43) | **(18.49)** |
| **ΔDeveloper** | -0.03 | -0.02 | **-0.05**** | -0.02 | -0.03 | 0.02 |
| | (-0.87) | (-0.67) | **(-2.98)** | (-0.75) | (-0.87) | (0.94) |
| N | 558 | 558 | 641 | 641 | 558 | 624 |
| Adj R-sq | 0.10 | 0.11 | 0.08 | 0.08 | 0.10 | 0.44 |
| **Panel B: Utilization** | | | | | | |
| | (1) ΔMktC_F | (2) ΔMktC_C | (3) ΔRevenue | (4) ΔTVL | (5) ΔCOMP | (6) ΔCOMP holder |
| Utilization | 0.00 | 0.01 | 0.00 | -0.01 | 0.00 | 0.01 |
| | (0.20) | (0.30) | (0.41) | (-0.76) | (0.20) | (0.82) |
| Hack | 0.00 | 0.00 | 0.00 | 0.00 | 0.00 | 0.00 |
| | (0.14) | (0.40) | (-0.64) | (0.42) | (0.14) | (0.64) |
| **ΔOutstanding loan** | **-0.08*** | -0.06 | 0.01 | -0.05 | **-0.08*** | -0.01 |
| | **(-1.82)** | (-1.57) | (0.27) | (-1.42) | **(-1.82)** | (-0.44) |
| **ΔOutstanding deposit** | **0.44**** | **0.43**** | **0.23**** | **0.35**** | **0.44**** | **0.07*** |
| | **(7.58)** | **(7.90)** | **(6.42)** | **(7.27)** | **(7.58)** | **(1.81)** |
| **Loan vol usd** | **-0.10**** | **-0.09*** | **-0.06*** | 0.03 | **-0.10**** | **0.10**** |
| | **(-1.96)** | **(-1.84)** | **(-1.68)** | (0.70) | **(-1.96)** | **(2.84)** |
| **Liquidation usd** | 0.09 | **0.12*** | -0.03 | 0.01 | 0.09 | 0.03 |
| | (1.20) | **(1.79)** | (-0.63) | (0.09) | (1.20) | (0.75) |
| **Active user** | 0.01 | 0.01 | 0.01 | 0.01 | 0.01 | **0.33**** |
| | (0.36) | (0.32) | (0.34) | (0.70) | (0.36) | **(20.83)** |
| **ΔDeveloper** | -0.02 | -0.02 | **-0.05**** | -0.02 | -0.02 | 0.02 |
| | (-0.77) | (-0.57) | **(-3.00)** | (-0.72) | (-0.77) | (1.26) |
| N | 558 | 558 | 641 | 641 | 558 | 624 |
| Adj R-sq | 0.09 | 0.10 | 0.08 | 0.08 | 0.09 | 0.43 |
| **Panel C: Repeat deposit ratio** | | | | | | |
| | (1) ΔMktC_F | (2) ΔMktC_C | (3) ΔRevenue | (4) ΔTVL | (5) ΔCOMP | (6) ΔCOMP holder |
| Repeat deposit ratio | 0.00 | 0.00 | 0.00 | -0.01 | 0.00 | -0.02 |
| | (0.13) | (0.11) | (-0.09) | (-0.39) | (0.13) | (-1.56) |
| Hack | 0.00 | 0.00 | 0.00 | 0.00 | 0.00 | 0.00 |
| | (0.15) | (0.43) | (-0.61) | (0.39) | (0.15) | (0.81) |
| **ΔOutstanding loan** | **-0.08*** | -0.06 | 0.01 | -0.05 | **-0.08*** | -0.01 |
| | **(-1.82)** | (-1.57) | (0.27) | (-1.41) | **(-1.82)** | (-0.40) |
| **ΔOutstanding deposit** | **0.44**** | **0.43**** | **0.23**** | **0.36**** | **0.44**** | **0.06*** |
| | **(7.60)** | **(7.91)** | **(6.41)** | **(7.41)** | **(7.60)** | **(1.74)** |
| **Loan vol usd** | **-0.10**** | **-0.09*** | **-0.05*** | 0.02 | **-0.10**** | **0.10**** |
| | **(-1.98)** | **(-1.82)** | **(-1.63)** | (0.47) | **(-1.98)** | **(3.25)** |
| **Liquidation usd** | 0.09 | **0.12*** | -0.03 | 0.01 | 0.09 | 0.03 |
| | (1.19) | **(1.78)** | (-0.65) | (0.13) | (1.19) | (0.70) |
| **Active user** | 0.01 | 0.01 | 0.00 | 0.02 | 0.01 | **0.33**** |
| | (0.34) | (0.28) | (0.29) | (0.78) | (0.34) | **(20.85)** |
| **ΔDeveloper** | -0.02 | -0.02 | **-0.05**** | -0.02 | -0.02 | 0.02 |
| | (-0.78) | (-0.58) | **(-3.02)** | (-0.69) | (-0.78) | (1.23) |
| N | 558 | 558 | 641 | 641 | 558 | 624 |
| Adj R-sq | 0.09 | 0.10 | 0.08 | 0.08 | 0.09 | 0.43 |
| **Panel D: Repeat loan ratio** | | | | | | |
| | (1) ΔMktC_F | (2) ΔMktC_C | (3) ΔRevenue | (4) ΔTVL | (5) ΔCOMP | (6) ΔCOMP holder |
| Repeat loan ratio | 0.03 | 0.02 | 0.00 | 0.01 | 0.03 | 0.01 |
| | (1.23) | (0.94) | (0.08) | (0.48) | (1.23) | (1.22) |
| Hack | 0.00 | 0.00 | 0.00 | 0.00 | 0.00 | 0.00 |
| | (0.07) | (0.36) | (-0.61) | (0.36) | (0.07) | (0.64) |

|  | (1) | (2) | (3) | (4) | (5) | (6) |
|---|---|---|---|---|---|---|
| ΔOutstanding loan | **-0.08*** | -0.06 | 0.01 | -0.05 | **-0.08*** | -0.01 |
|  | **(-1.83)** | (-1.57) | (0.27) | (-1.42) | **(-1.83)** | (-0.45) |
| ΔOutstanding deposit | **0.43*** | **0.42*** | **0.23*** | **0.36*** | **0.43*** | 0.06 |
|  | **(7.41)** | **(7.75)** | **(6.38)** | **(7.34)** | **(7.41)** | (1.60) |
| Loan vol usd | **-0.10*** | **-0.08*** | **-0.05*** | 0.02 | **-0.10*** | **0.11*** |
|  | **(-1.87)** | **(-1.74)** | **(-1.63)** | (0.50) | **(-1.88)** | **(3.36)** |
| Liquidation usd | 0.09 | **0.12*** | -0.03 | 0.01 | 0.09 | 0.03 |
|  | (1.24) | **(1.82)** | (-0.65) | (0.15) | (1.24) | (0.76) |
| Active user | 0.01 | 0.01 | 0.00 | 0.02 | 0.01 | **0.33*** |
|  | (0.38) | (0.30) | (0.30) | (0.79) | (0.37) | **(20.89)** |
| ΔDeveloper | -0.02 | -0.02 | **-0.05*** | -0.02 | -0.02 | 0.02 |
|  | (-0.77) | (-0.58) | **(-3.02)** | (-0.69) | (-0.77) | (1.23) |
| N | 558 | 558 | 641 | 641 | 558 | 624 |
| Adj R-sq | 0.09 | 0.10 | 0.08 | 0.08 | 0.09 | 0.43 |

Note: This table reports regression results for influence of liquidity risks and DeFi hacks on Compound protocol. In columns (1) – (6) of each panel, the dependent variable is $\Delta MktC\_F, \Delta MktC\_C, \Delta Revenue, \Delta TVL, \Delta COMP$, and $\Delta COMP\ holder$ respectively. T-statistics are reported in parentheses. *, **, and *** denote significance levels at the 10%, 5%, and 1% levels based on the standard t-statistics.

**Table 11. Introduction to mainstream cryptocurrencies traded in Aave**

| Token | Description |
|---|---|
| **ETH** | Ether (ETH) is the native token of Ethereum blockchain. |
| **WBTC** | Wrapped Bitcoin (WBTC) mirrors the value of Bitcoin so that can be traded on Ethereum blockchain. |
| **USDC** | USD Coin (USDC) is a fiat-backed stablecoin that aims to maintain the 1:1 peg with the U.S. dollar. The backed assets are held in accounts with US regulated financial institutions. |
| **USDT** | Tether (USDT) is a stablecoin pegged to the U.S. dollar and backed by Tether's collateral assets. The backed assets are held in accounts with US regulated financial institutions. |
| **DAI** | Dai (DAI) is a stablecoin pegged to the U.S. dollar and collateralized by tokens. |
| **TUSD** | TrueUSD (TUSD) is a fiat-backed stablecoin that aims to maintain the 1:1 peg with the U.S. dollar. |
| **sUSD** | sUSD is a synthetic stablecoin issued by Synthetix protocol and tacks the value of the U.S. dollar. |
| **BUSD** | Binance USD (BUSD) is a fiat-backed stablecoin that aims to maintain the 1:1 peg with the U.S. dollar. |
| **FEI** | Fei USD (FEI) is a stablecoin backed 1:1 by DAI stablecoin. |
| **FRAX** | Frax (Frax) is a stablecoin backed by both collateral assets and algorithms. |
| **AAVE** | Aave (AAVE) is the governance token of Aave protocol. |
| **CRV** | Curve (CRV) is the governance token of Curve finance. |
| **SNX** | Synthetix (SNX) is used as both collateral and governance token in Synthetix protocol. |
| **YFI** | Yearn.finance (YFI) is the governance token of Yearn finance. |
| **LINK** | Chainlink (LINK), issued by Chainlink, is a token aiming to incentivize computers to provide reliable real-world data. |
| **stETH** | Staked Ether (stETH) is a cryptocurrency token that aims to represent an Ethereum token that is "staked" or deposited to support blockchain operations. |

Note: This table gives brief introduction to mainstream cryptocurrencies traded in Aave v1 and v2 (from Dec 1st, 2020 to Jan 31st, 2023).

**Table 12. Measurements of liquidity risks for mainstream cryptocurrencies**

|  | Liquidity | Utilization | Repeat deposit ratio | Repeat loan ratio |
|---|---|---|---|---|
| **Mean** | 3534666107.51 | 0.36 | 0.86 | 0.86 |
| **Median** | 3918117152.00 | 0.31 | 0.92 | 0.93 |
| **Maximum** | 8159906995.00 | 0.75 | 1.00 | 1.00 |
| **Minimum** | 0.00 | 0 | 0.00 | 0.00 |
| **Std** | 2893608151.27 | 0.15 | 0.17 | 0.17 |

Note: This table reports descriptive statistics of liquidity risk measurements based on the datasets for mainstream cryptocurrencies traded in Aave V1 and V2 lending pool (from Dec 16th, 2019 to Jan 31st, 2023).

**Table 13. The effects of liquidity risks caused by mainstream cryptocurrencies on Aave protocol**

| Panel A: Liquidity | | | | | | |
|---|---|---|---|---|---|---|
|  | (1) | (2) | **(3)** | (4) | (5) | **(6)** |
|  | ΔMktC_F | ΔMktC_C | **Revenue** | ΔTVL | ΔAAVE | **ΔAAVE holder** |
| **Liquidity** | -0.01 | -0.01 | **0.15*** | 0.00 | -0.01 | **-0.11*** |
|  | (-1.09) | (-1.08) | **(9.25)** | (-0.44) | (-1.09) | **(-11.25)** |

|  | V2 | -0.01 | -0.01 | 0.06 | 1.85e+10 | -0.01 | -0.01 |
|---|---|---|---|---|---|---|---|
|  |  | (-0.15) | (-0.14) | (0.63) | (-0.03) | (-0.15) | (-0.10) |
| **Hack** |  | 0.00 | 0.00 | **0.06*** | 0.01 | 0.00 | **-0.02*** |
|  |  | (0.09) | (0.16) | **(5.10)** | (1.14) | (0.09) | **(-3.08)** |
| **ΔOutstanding loan** |  | 0.06 | 0.07 | **-0.26*** | **0.16**** | 0.06 | -0.10 |
|  |  | (0.93) | (1.03) | **(-1.80)** | **(2.25)** | (0.93) | (-1.09) |
| **ΔOutstanding deposit** |  | 0.02 | 0.01 | **-0.35**** | -0.02 | 0.02 | 0.02 |
|  |  | (0.27) | (0.18) | **(-2.80)** | (-0.24) | (0.27) | (0.31) |
| **Deposits vol usd** |  | -0.04 | -0.04 | **0.42**** | -0.05 | -0.04 | **-0.12*** |
|  |  | (-1.21) | (-1.16) | **(6.76)** | (-1.62) | (-1.22) | **(-3.16)** |
| **Loan vol usd** |  | 0.01 | 0.01 | 0.09 | **0.06*** | 0.01 | **0.11*** |
|  |  | (0.33) | (0.15) | (1.27) | **(1.64)** | (0.33) | **(2.59)** |
| **Liquidation usd** |  | 0.01 | -0.06 | 0.03 | **-0.14**** | -0.03 | 0.03 |
|  |  | (0.33) | (-1.18) | (0.26) | **(-2.52)** | (-0.55) | (0.43) |
| **ΔActive user** |  | 0.03 | 0.04 | 0.02 | **0.05*** | 0.03 | 0.02 |
|  |  | (1.39) | (1.49) | (0.38) | **(1.92)** | (1.39) | (0.65) |
| ΔDeveloper |  | -0.03 | -0.03 | -0.03 | -0.02 | -0.03 | 0.01 |
|  |  | (-1.44) | (-1.26) | (-0.56) | (-0.68) | (-1.43) | (0.53) |
| N |  | 792 | 792 | 792 | 789 | 792 | 792 |
| Adj R-sq |  | 0.00 | 0.00 | 0.23 | 0.03 | 0.00 | 0.15 |

**Panel B: Utilization**

|  | (1) | (2) | **(3)** | (4) | (5) | (6) |
|---|---|---|---|---|---|---|
|  | ΔMktC_F | ΔMktC_C | **Revenue** | ΔTVL | ΔAAVE | ΔAAVE holder |
| **Utilization** | -0.01 | 0.00 | **0.41*** | 0.01 | -0.01 | -0.01 |
|  | (-0.44) | (-0.34) | **(17.03)** | (1.02) | (-0.44) | (-0.48) |
| V2 | -0.01 | -0.01 | 0.06 | -6.65e+10 | -0.01 | -0.06 |
|  | (-0.22) | (-0.21) | (0.72) | (-0.09) | (-0.22) | (-0.92) |
| **Hack** | 0.00 | 0.00 | **0.04*** | 0.01 | 0.00 | **-0.02*** |
|  | (0.12) | (0.18) | **(3.80)** | (1.01) | (0.12) | **(-2.98)** |
| **ΔOutstanding loan** | 0.07 | 0.08 | **-0.57*** | **0.16**** | 0.07 | 0.03 |
|  | (1.08) | (1.19) | **(-4.42)** | **(2.24)** | (1.09) | (0.27) |
| ΔOutstanding deposit | 0.01 | 0.01 | -0.16 | -0.01 | 0.01 | -0.04 |
|  | (0.17) | (0.09) | (-1.38) | (-0.21) | (0.17) | (-0.45) |
| **Deposits vol usd** | -0.03 | -0.03 | 0.22 | **-0.06*** | -0.03 | **-0.12*** |
|  | (-1.11) | (-1.07) | (3.86) | **(-1.80)** | (-1.11) | **(-2.86)** |
| **Loan vol usd** | 0.01 | 0.00 | **0.20**** | **0.06*** | 0.01 | 0.05 |
|  | (0.18) | (0.01) | **(3.20)** | **(1.63)** | (0.18) | (1.10) |
| **Liquidation usd** | -0.02 | -0.06 | -0.07 | **-0.14**** | -0.02 | 0.08 |
|  | (-0.48) | (-1.11) | (-0.68) | **(-2.52)** | (-0.48) | (1.09) |
| **ΔActive user** | 0.03 | 0.04 | 0.02 | **0.05*** | 0.03 | 0.02 |
|  | (1.39) | (1.50) | (0.36) | **(1.93)** | (1.39) | (0.68) |
| ΔDeveloper | -0.03 | -0.03 | -0.03 | -0.02 | -0.03 | 0.02 |
|  | (-1.42) | (-1.25) | (-0.66) | (-0.66) | (-1.42) | (0.63) |
| N | 792 | 792 | 792 | 789 | 792 | 792 |
| Adj R-sq | 0.53 | 0.00 | 0.38 | 0.03 | 0.00 | 0.01 |

**Panel C: Repeat deposit ratio**

|  | (1) | (2) | (3) | (4) | (5) | (6) |
|---|---|---|---|---|---|---|
|  | ΔMktC_F | ΔMktC_C | Revenue | ΔTVL | ΔAAVE | ΔAAVE holder |
| Repeat deposit ratio | 0.01 | 0.00 | 0.04 | 0.02 | 0.01 | 0.01 |
|  | (0.44) | (0.28) | (1.00) | (0.91) | (0.44) | (0.65) |
| V2 | -0.01 | -0.01 | 0.12 | -1.57e+11 | -0.01 | -0.06 |
|  | (-0.26) | (-0.24) | (1.26) | (-0.21) | (-0.26) | (-0.98) |
| **Hack** | 0.00 | 0.00 | **0.06*** | 0.01 | 0.00 | **-0.02*** |
|  | (0.06) | (0.13) | **(4.98)** | (1.11) | (0.06) | **(-3.08)** |
| **ΔOutstanding loan** | 0.07 | 0.08 | **-0.42*** | **0.17**** | 0.07 | 0.02 |
|  | (1.06) | (1.17) | **(-2.77)** | **(2.31)** | (1.06) | (0.24) |
| **ΔOutstanding deposit** | 0.01 | 0.01 | **-0.27**** | -0.02 | 0.01 | -0.03 |
|  | (0.20) | (0.11) | **(-2.05)** | (-0.26) | (0.20) | (-0.42) |
| **Deposits vol usd** | -0.04 | -0.04 | **0.42*** | -0.05 | -0.04 | **-0.12*** |
|  | (-1.22) | (-1.17) | **(6.50)** | (-1.62) | (-1.22) | **(-3.02)** |
| **Loan vol usd** | 0.01 | 0.00 | **0.18**** | **0.06*** | 0.01 | 0.05 |
|  | (0.24) | (0.04) | **(2.40)** | **(1.68)** | (0.24) | (1.18) |
| **Liquidation usd** | -0.02 | -0.06 | -0.03 | **-0.14**** | -0.02 | 0.08 |
|  | (-0.47) | (-1.10) | (-0.29) | **(-2.46)** | (-0.47) | (1.11) |
| **ΔActive user** | 0.03 | 0.04 | 0.01 | **0.05*** | 0.03 | 0.02 |
|  | (1.37) | (1.48) | (0.24) | **(1.87)** | (1.37) | (0.65) |
| ΔDeveloper | -0.03 | -0.03 | -0.04 | -0.02 | -0.03 | 0.02 |
|  | (-1.45) | (-1.27) | (-0.76) | (-0.75) | (-1.45) | (0.57) |
| N | 792 | 792 | 792 | 789 | 792 | 792 |
| Adj R-sq | 0.00 | 0.00 | 0.15 | 0.02 | 0.00 | 0.01 |

**Panel D: Repeat loan ratio**

|  | **(1)** | **(2)** | **(3)** | (4) | **(5)** | **(6)** |
|---|---|---|---|---|---|---|
|  | **ΔMktC_F** | **ΔMktC_C** | **Revenue** | ΔTVL | **ΔAAVE** | **ΔAAVE holder** |
| **Repeat loan ratio** | **0.03**** | **0.03*** | **-0.08**** | 0.01 | **0.03**** | **0.04**** |

|  | (2.14) | (1.91) | (-2.24) | (0.63) | (2.14) | (1.93) |
|---|---|---|---|---|---|---|
| V2 | -0.01 | -0.01 | 0.13 | -3.03e+10 | -0.01 | -0.06 |
|  | (-0.29) | (-0.27) | (1.38) | (-0.04) | (-0.29) | (-0.99) |
| **Hack** | 0.00 | 0.00 | **0.07***| 0.01 | 0.00 | **-0.02*** |
|  | (0.07) | (0.14) | **(5.04)** | (1.13) | (0.07) | **(-3.07)** |
| **ΔOutstanding loan** | 0.07 | 0.08 | **-0.42*** | **0.17**| 0.07 | 0.02 |
|  | (1.06) | (1.17) | **(-2.78)** | **(2.32)** | (1.06) | (0.23) |
| **ΔOutstanding deposit** | 0.01 | 0.00 | **-0.26**| -0.02 | 0.01 | -0.04 |
|  | (0.12) | (0.03) | **(-1.98)** | (-0.29) | (0.12) | (-0.50) |
| **Deposits vol usd** | -0.03 | -0.03 | **0.41*** | -0.05 | -0.03 | **-0.12*** |
|  | (-1.05) | (-1.01) | **(6.31)** | (-1.57) | (-1.05) | **(-2.86)** |
| **Loan vol usd** | 0.01 | 0.01 | **0.15**| **0.06***| 0.01 | 0.06 |
|  | (0.39) | (0.19) | **(2.10)** | **(1.65)** | (0.39) | (1.29) |
| **Liquidation usd** | -0.02 | -0.05 | -0.06 | **-0.14**| -0.02 | 0.09 |
|  | (-0.30) | (-0.95) | (-0.53) | **(-2.44)** | (-0.30) | (1.25) |
| **ΔActive user** | 0.03 | 0.04 | 0.01 | **0.05**| 0.03 | 0.02 |
|  | (1.41) | (1.51) | (0.28) | **(1.93)** | (1.41) | (0.70) |
| ΔDeveloper | -0.03 | -0.03 | -0.02 | -0.02 | -0.03 | 0.01 |
|  | (-1.59) | (-1.40) | (-0.50) | (-0.72) | (-1.59) | (0.48) |
| N | 792 | 792 | 792 | 789 | 792 | 792 |
| Adj R-sq | 0.00 | 0.01 | 0.16 | 0.03 | 0.00 | 792 |

Note: This table reports regression results for influence of liquidity risks caused by mainstream cryptocurrencies on Aave protocol. In columns (1) – (6) of each panel, the dependent variable is $\Delta MktC\_F, \Delta MktC\_C, Revenue, \Delta TVL, \Delta AAVE$, and $\Delta AAVE\ holder$ respectively. T-statistics are reported in parentheses. *, **, and *** denote significance levels at the 10%, 5%, and 1% levels based on the standard t-statistics.

**Table 14. The effects of liquidity risks caused by mainstream cryptocurrencies on Compound protocol**

| **Panel A: Liquidity** | | | | | | |
|---|---|---|---|---|---|---|
|  | **(1)** | **(2)** | (3) | (4) | **(5)** | **(6)** |
|  | **ΔMktC_F** | **ΔMktC_C** | ΔRevenue | ΔTVL | **ΔCOMP** | **ΔCOMP holder** |
| **Liquidity** | **-0.02*** | **-0.02*** | 0.00 | -0.01 | **-0.02*** | **-0.03*** |
|  | **(-1.80)** | **(-1.86)** | (0.74) | (-1.25) | **(-1.80)** | **(-4.04)** |
| V3 | 0.00 | 0.00 | 0.00 | 0.00 | 0.00 | 0.01 |
|  | (0.17) | (0.23) | (-0.21) | (0.09) | (0.17) | (1.10) |
| Hack | 0.00 | 0.00 | 0.00 | 0.00 | 0.00 | 0.00 |
|  | (0.18) | (0.46) | (-0.59) | (0.33) | (0.18) | (0.69) |
| **ΔOutstanding loan** | **-0.07*** | -0.06 | 0.01 | -0.05 | **-0.07*** | -0.01 |
|  | **(-1.79)** | (-1.53) | (0.26) | (-1.40) | **(-1.79)** | (-0.37) |
| **ΔOutstanding deposit** | **0.43*** | **0.42*** | **0.23*** | **0.35*** | **0.43*** | 0.04 |
|  | **(7.37)** | **(7.67)** | **(6.45)** | **(7.18)** | **(7.37)** | (1.19) |
| **Loan vol usd** | **-0.11**| **-0.10**| **-0.05*** | 0.02 | **-0.11**| **0.11*** |
|  | **(-2.15)** | **(-2.00)** | **(-1.64)** | (0.49) | **(-2.15)** | **(3.38)** |
| Liquidation usd | 0.07 | 0.11 | -0.02 | 0.00 | 0.07 | 0.01 |
|  | (0.98) | (1.56) | (-0.57) | (-0.01) | (0.98) | (0.28) |
| **Active user** | -0.01 | -0.01 | 0.01 | 0.01 | -0.01 | **0.31*** |
|  | (-0.44) | (-0.51) | (0.49) | (0.35) | (-0.44) | **(18.64)** |
| ΔDeveloper | -0.03 | -0.02 | **-0.05**| -0.02 | -0.03 | 0.02 |
|  | (-0.87) | (-0.67) | **(-2.98)** | (-0.76) | (-0.87) | (0.96) |
| N | 558 | 558 | 641 | 641 | 558 | 624 |
| Adj R-sq | 0.09 | 0.11 | 0.08 | 0.08 | 0.09 | 0.44 |
| **Panel B: Utilization** | | | | | | |
|  | (1) | (2) | (3) | (4) | (5) | (6) |
|  | ΔMktC_F | ΔMktC_C | ΔRevenue | ΔTVL | ΔCOMP | ΔCOMP holder |
| Utilization | 0.01 | 0.01 | 0.00 | -0.01 | 0.01 | 0.00 |
|  | (0.33) | (0.34) | (0.19) | (-0.83) | (0.33) | (-0.13) |
| V3 | 0.00 | 0.00 | 0.00 | 0.00 | 0.00 | 0.00 |
|  | (-0.02) | (0.03) | (0.00) | (-0.38) | (-0.02) | (0.21) |
| Hack | 0.00 | 0.00 | 0.00 | 0.00 | 0.00 | 0.00 |
|  | (0.13) | (0.40) | (-0.62) | (0.44) | (0.13) | (0.71) |
| **ΔOutstanding loan** | **-0.08*** | -0.06 | 0.01 | -0.05 | **-0.08*** | -0.01 |
|  | **(-1.82)** | (-1.56) | (0.27) | (-1.42) | **(-1.82)** | (-0.44) |
| **ΔOutstanding deposit** | **0.44*** | **0.43*** | **0.23*** | **0.35*** | **0.44*** | **0.06*** |
|  | **(7.58)** | **(7.89)** | **(6.38)** | **(7.26)** | **(7.58)** | **(1.67)** |
| **Loan vol usd** | **-0.11**| **-0.09*** | -0.05 | 0.03 | **-0.11**| **0.11*** |
|  | **(-2.00)** | **(-1.85)** | (-1.61) | (0.68) | **(-2.00)** | **(3.19)** |
| **Liquidation usd** | 0.09 | **0.12***| -0.03 | 0.00 | 0.09 | 0.03 |
|  | (1.20) | **(1.79)** | (-0.64) | (0.08) | (1.20) | (0.70) |
| **Active user** | 0.01 | 0.01 | 0.01 | 0.01 | 0.01 | **0.33*** |
|  | (0.38) | (0.34) | (0.32) | (0.54) | (0.38) | **(20.08)** |
| **ΔDeveloper** | -0.02 | -0.02 | **-0.05****| -0.02 | -0.02 | 0.02 |

| | (-0.76) | (-0.56) | **(-3.00)** | (-0.75) | (-0.76) | (1.21) |
| N | 558 | 558 | 641 | 641 | 548 | 624 |
| Adj R-sq | 0.09 | 0.10 | 0.07 | 0.08 | 0.09 | 0.43 |
| **Panel C: Repeat deposit ratio** | | | | | | |
| | (1) | (2) | (3) | (4) | (5) | **(6)** |
| | ΔMktC_F | ΔMktC_C | ΔRevenue | ΔTVL | ΔCOMP | **ΔCOMP holder** |
| **Repeat deposit ratio** | 0.01 | 0.01 | 0.00 | -0.01 | 0.01 | **-0.02*** |
| | (0.25) | (0.25) | (-0.15) | (-0.35) | (0.25) | **(-1.71)** |
| V3 | 0.00 | 0.00 | 0.00 | 0.00 | 0.00 | 0.00 |
| | (-0.15) | (-0.10) | (-0.03) | (-0.15) | (-0.15) | (0.47) |
| Hack | 0.00 | 0.00 | 0.00 | 0.00 | 0.00 | 0.00 |
| | (0.15) | (0.42) | (-0.60) | (0.39) | (0.15) | (0.80) |
| **ΔOutstanding loan** | **-0.08*** | -0.06 | 0.01 | -0.05 | **-0.08*** | -0.01 |
| | **(-1.82)** | (-1.57) | (0.27) | (-1.41) | **(-1.82)** | (-0.40) |
| **ΔOutstanding deposit** | **0.44*** | **0.43*** | **0.23*** | **0.36*** | **0.44*** | **0.06*** |
| | **(7.58)** | **(7.89)** | **(6.40)** | **(7.41)** | **(7.58)** | **(1.72)** |
| **Loan vol usd** | **-0.10*** | **-0.09*** | -0.05 | 0.02 | **-0.10*** | **0.11*** |
| | **(-1.97)** | **(-1.81)** | (-1.62) | (0.45) | **(-1.97)** | **(3.28)** |
| **Liquidation usd** | 0.09 | **0.12*** | -0.03 | 0.01 | 0.09 | 0.03 |
| | (1.19) | **(1.78)** | (-0.65) | (0.13) | (1.19) | (0.74) |
| **Active user** | 0.01 | 0.01 | 0.00 | 0.02 | 0.01 | **0.33*** |
| | (0.31) | (0.26) | (0.28) | (0.75) | (0.31) | **(20.67)** |
| **ΔDeveloper** | -0.02 | -0.02 | **-0.05*** | -0.02 | -0.02 | 0.02 |
| | (-0.78) | (-0.58) | **(-3.02)** | (-0.69) | (-0.78) | (1.23) |
| N | 558 | 558 | 641 | 641 | 558 | 624 |
| Adj R-sq | 0.09 | 0.10 | 0.07 | 0.08 | 0.09 | 0.43 |
| **Panel D: Repeat loan ratio** | | | | | | |
| | (1) | (2) | (3) | (4) | (5) | (6) |
| | ΔMktC_F | ΔMktC_C | ΔRevenue | ΔTVL | ΔCOMP | ΔCOMP holder |
| Repeat loan ratio | 0.03 | 0.02 | -0.01 | 0.01 | 0.03 | 0.01 |
| | (1.23) | (0.98) | (-0.91) | (0.69) | (1.23) | (0.99) |
| V3 | 0.00 | 0.00 | 0.00 | 0.00 | 0.00 | 0.00 |
| | (-0.27) | (-0.19) | (0.08) | (-0.29) | (-0.27) | (0.11) |
| Hack | 0.00 | 0.00 | 0.00 | 0.00 | 0.00 | 0.00 |
| | (0.06) | (0.35) | (-0.59) | (0.36) | (0.06) | (0.64) |
| **ΔOutstanding loan** | **-0.08*** | -0.06 | 0.01 | -0.05 | **-0.08*** | -0.01 |
| | **(-1.84)** | (-1.58) | (0.28) | (-1.43) | **(-1.84)** | (-0.45) |
| **ΔOutstanding deposit** | **0.43*** | **0.42*** | **0.23*** | **0.35*** | **0.43*** | 0.06 |
| | **(7.41)** | **(7.74)** | **(6.46)** | **(7.33)** | **(7.41)** | (1.61) |
| **Loan vol usd** | **-0.10*** | **-0.08*** | **-0.05*** | 0.02 | **-0.10*** | **0.11*** |
| | **(-1.89)** | **(-1.75)** | **(-1.64)** | (0.46) | **(-1.89)** | **(3.33)** |
| **Liquidation usd** | 0.09 | **0.12*** | -0.03 | 0.01 | 0.09 | 0.03 |
| | (1.23) | **(1.81)** | (-0.68) | (0.15) | (1.23) | (0.75) |
| **Active user** | 0.01 | 0.01 | 0.00 | 0.02 | 0.01 | **0.33*** |
| | (0.33) | (0.28) | (0.30) | (0.74) | (0.33) | **(20.64)** |
| **ΔDeveloper** | -0.02 | -0.02 | **-0.05*** | -0.02 | -0.02 | 0.02 |
| | (-0.79) | (-0.59) | **(-2.99)** | (-0.72) | (-0.79) | (1.21) |
| N | 558 | 558 | 641 | 641 | 558 | 624 |
| Adj R-sq | 0.09 | 0.10 | 0.08 | 0.08 | 0.09 | 0.43 |

Note: This table reports regression results for influence of liquidity risks caused by mainstream cryptocurrencies on Compound protocol. In columns (1) – (6) of each panel, the dependent variable is $\Delta MktC\_F, \Delta MktC\_C, \Delta Revenue, \Delta TVL, \Delta COMP$, and $\Delta COMP\ holder$ respectively. T-statistics are reported in parentheses. *, **, and *** denote significance levels at the 10%, 5%, and 1% levels based on the standard t-statistics.

# Figures

**Figure 1. Pooled funds in lending protocols**

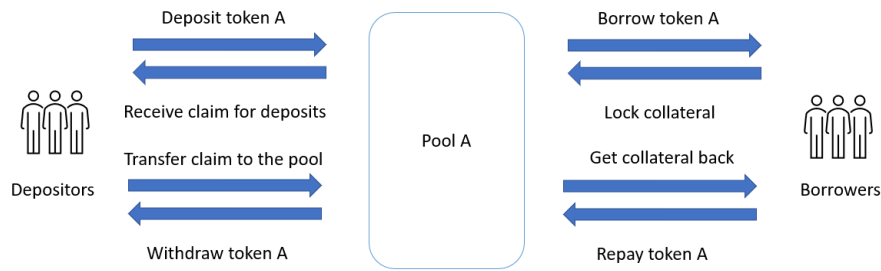

Note: This figure illustrates borrowing and lending in lending protocols. For each token, there will be a pool. Depositors can deposit their token and receive an amount of claim. When depositors want to withdraw their tokens, they need to transfer claim to the lending protocol. Borrowers need to lock collateral when requiring loans. When they successfully repay loans, the collateral can be returned.

**Figure 2. Liquidation in lending protocols**

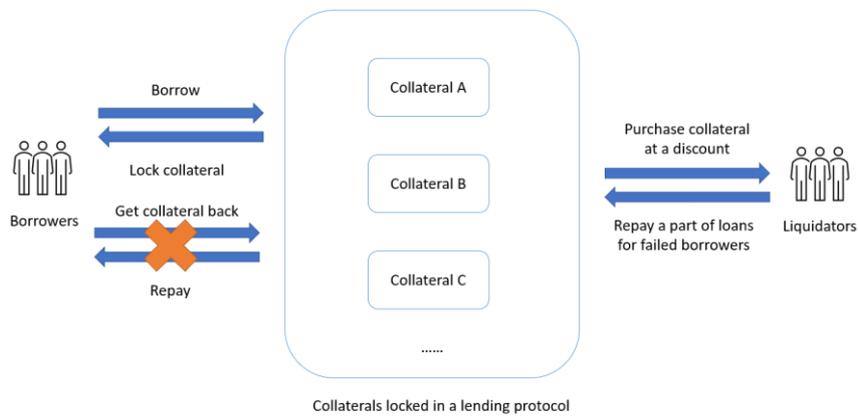

Note: This figure illustrates the liquidation in a lending pool. When borrowers fail to repay their loans, liquidators can participate in liquidation. Liquidators can repay a part of loans for failed borrowers. In return, liquidators can purchase collateral of failed borrowers at a discount.

**Figure 3. Outstanding debt and deposit in Aave protocol (Dec 16, 2019 – Jan 31, 2023)**

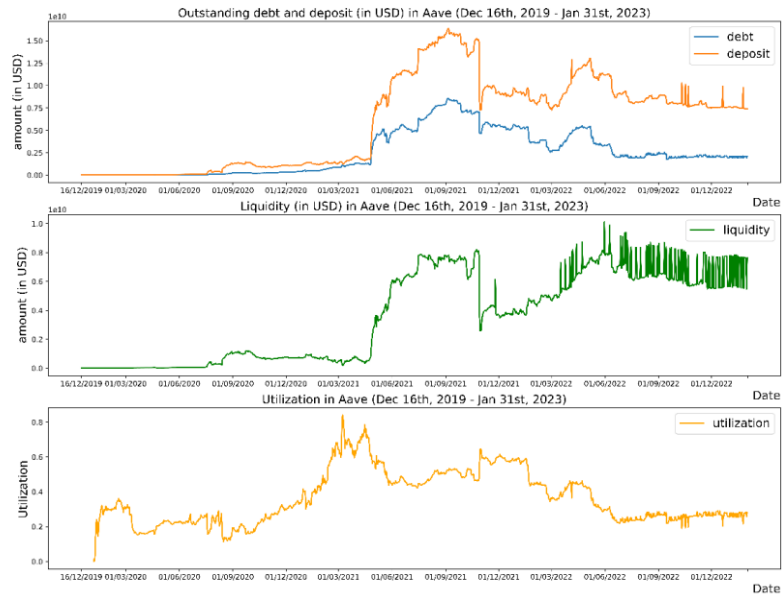

Note: This figure illustrates outstanding debt and deposit (in USD) in Aave protocol, and available liquidity (in USD) daily and utilization are also presented. In this figure, we consider all tokens traded in Aave.

**Figure 4. Deposit volume in Aave protocol (Dec 16, 2019 – Jan 31, 2023)**

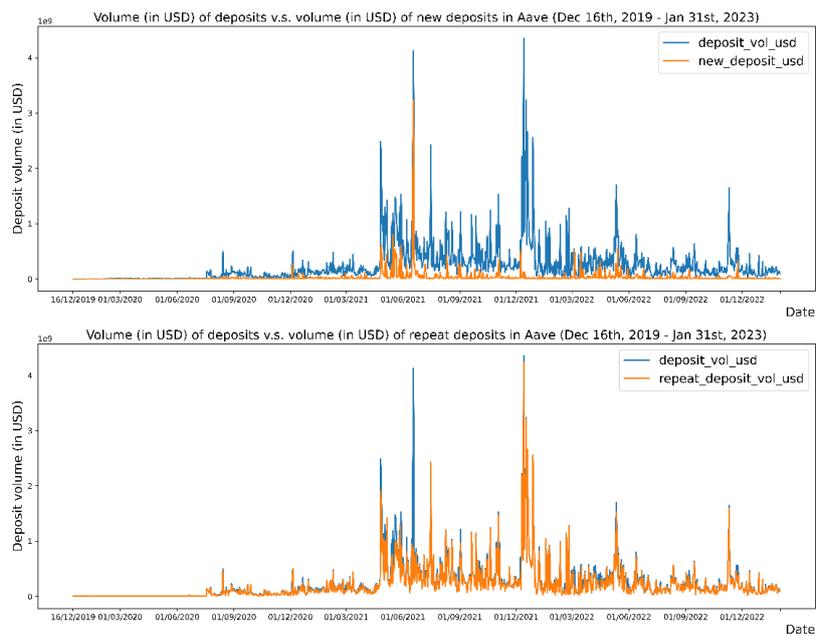

Note: This figure illustrates the volume (in USD) of deposits in Aave protocol, and we also show the volumes (in USD) of deposits from new depositors and repeat depositors.

**Figure 5. Loan volume in Aave protocol (Dec 16, 2019 – Jan 31, 2023)**

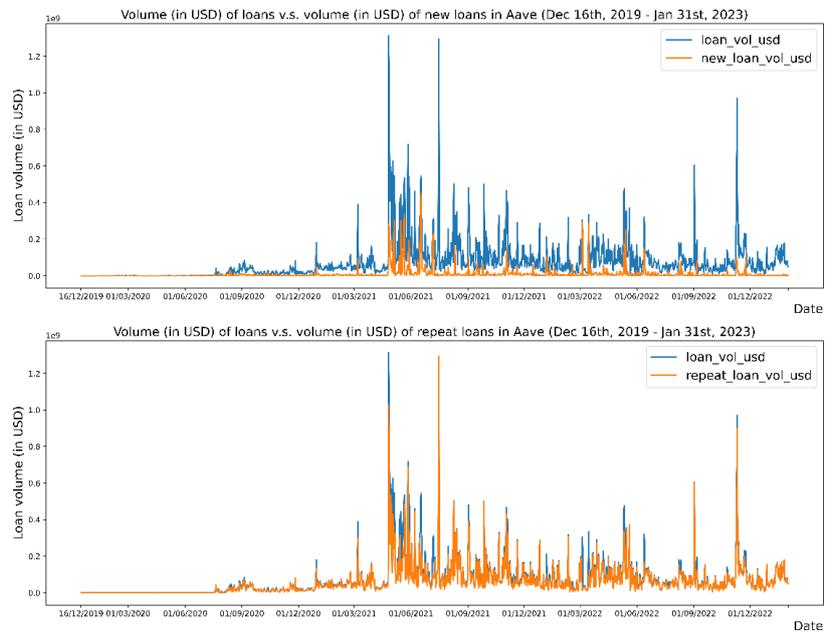

Note: This figure illustrates the volume (in USD) of loans in Aave protocol, and we also show the volumes (in USD) of loans from new borrowers and repeat borrowers.

**Figure 6. The proportion of repeat deposits and repeat loans (Dec 16, 2019 – Jan 31, 2023)**

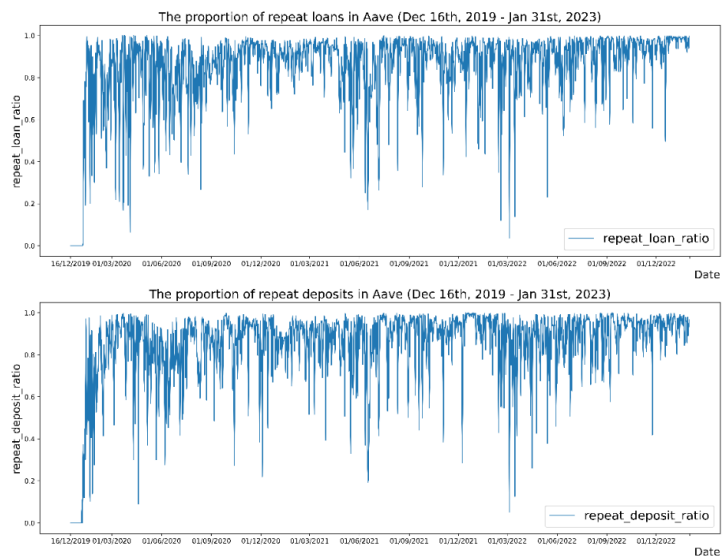

Note: This figure illustrates the proportion of repeat loans and deposits in Aave protocol, respectively.

**Figure 7. Outstanding debt and deposit of mainstream tokens traded in Aave protocol (Dec 16, 2019 – Jan 31, 2023)**

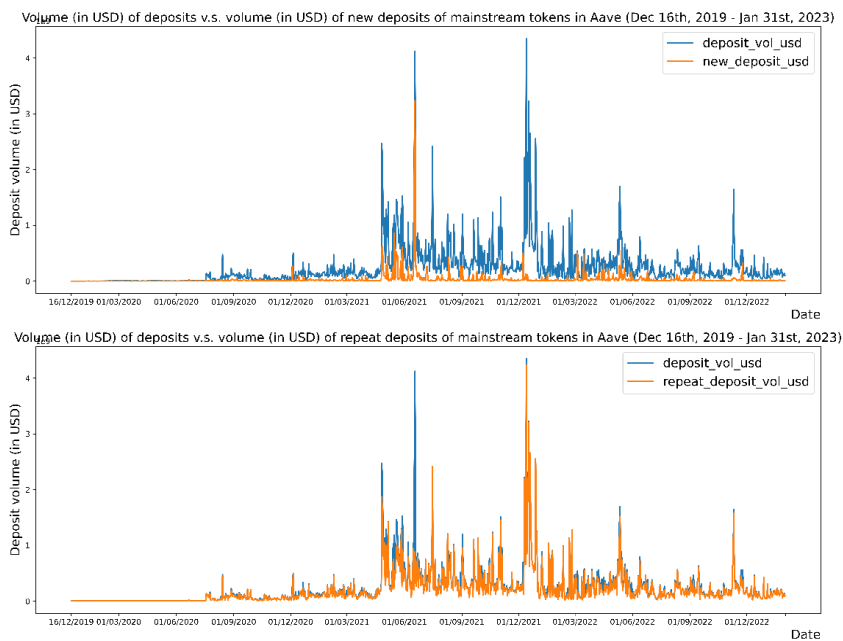

Note: This figure illustrates outstanding debt and deposit (in USD) of mainstream tokens traded in Aave protocol, and available liquidity (in USD) daily and utilization are also presented.

**Appendix**

This appendix section includes further technical information complementing the main text of this study. First, appendix A.1 presents details about loans and deposits in Aave. In appendix A.2, we explain the definitions of Compound-specific factors. Finally, appendix A.3 is about mainstream cryptocurrencies traded on Aave and their borrowing and lending activities on Aave protocol.

**A.1 Details about loans and deposits in Aave**

**Table A.1: Variables related to loans and deposits in Aave**

| Panel A: Loan details | |
|---|---|
| **Variable** | **Description** |
| **Borrower** | The number of borrowers daily |
| **Loan vol usd** | Daily volume (in USD) of Aave loans |
| **Loan cnt** | The count of Aave loans daily |
| **New borrower** | The number of new borrowers daily |
| **New loan vol usd** | Daily volume (in USD) of Aave loans initiated by new borrowers |
| **New loan cnt** | The count of Aave loans initiated by new borrowers daily |
| **Avg loan usd** | Daily volume (in USD) of Aave loans divided by the count of Aave loans daily |
| **Outstanding loan** | The value (in USD) of outstanding loans in Aave |
| **Liquidation usd** | The value (in USD) of collateral liquidated daily in Aave |
| **Repeat borrower** | The number of borrowers daily minus the number of new borrowers daily |
| **Repeat loan vol usd** | Daily volume (in USD) of Aave loans initiated by repeat borrowers |
| **Repeat loan cnt** | The count of Aave loans initiated by repeat borrowers daily |
| **Panel B: Deposit details** | |
| **Variable** | **Description** |
| **Depositor** | The number of depositors daily |
| **Deposit vol usd** | Daily volume (in USD) of Aave deposits |
| **Deposit cnt** | The count of Aave deposits daily |
| **New depositor** | The number of new depositors daily |
| **New deposit vol usd** | Daily volume (in USD) of Aave deposits from new depositors |
| **New deposit cnt** | The count of Aave deposits from new depositors daily |
| **Avg deposit usd** | Daily volume (in USD) of Aave deposits divided by the count of Aave deposits daily |
| **Outstanding deposit** | The value (in USD) of outstanding deposits in Aave |
| **Repeat depositor** | The number of depositors daily minus the number of new depositors daily |
| **Repeat deposit vol usd** | Daily volume (in USD) of Aave deposits from repeat depositors |
| **Repeat deposit cnt** | The count of Aave deposits from repeat depositors daily |

Note: This table introduces the variables related to details of loans and deposits in Aave protocol.

**Figure A.1: Depositors in Aave protocol (Dec 16, 2019 – Jan 31, 2023)**

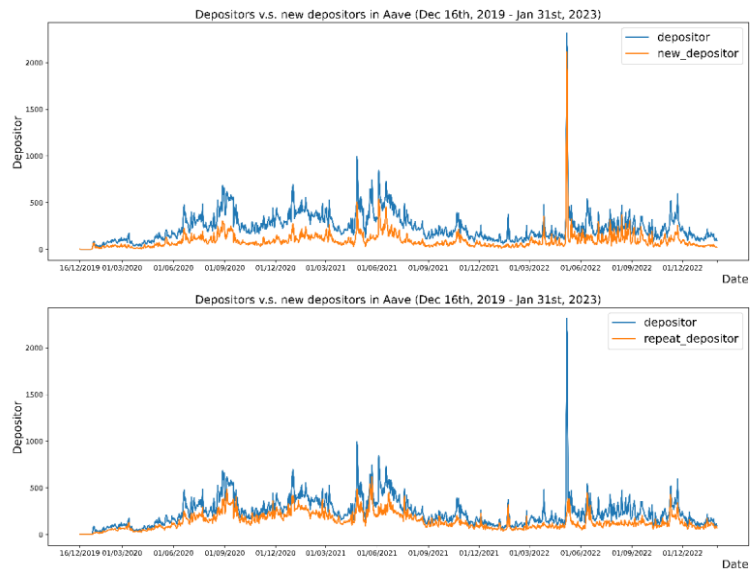

Note: This figure illustrates the number of depositors daily in Aave protocol, and we also show the number of new depositors and repeat depositors daily.

**Figure A.2: Borrowers in Aave protocol (Dec 16, 2019 – Jan 31, 2023)**

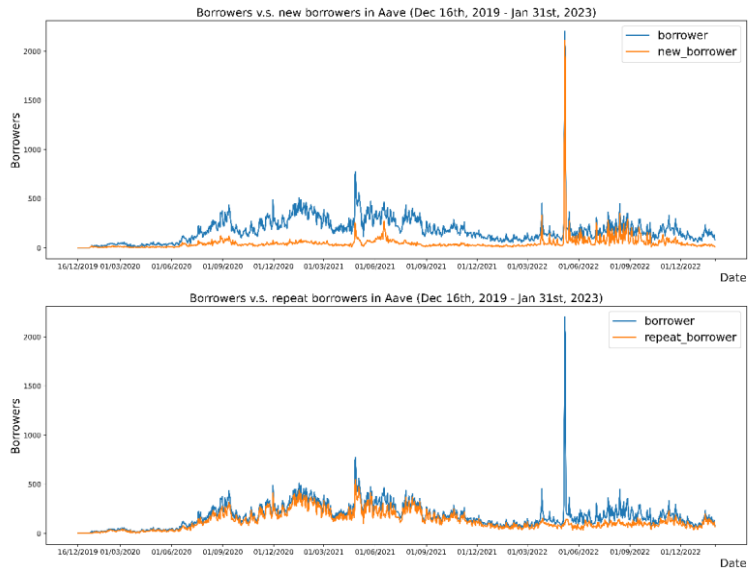

Note: This figure illustrates the number of borrowers daily in Aave protocol, and we also show the number of new borrowers and repeat borrowers daily.

## A.2 Cross-LP effects

**Table A.2: Compound protocol-specific factors**

| Factor | Definition |
|---|---|
| **MktC_F** | Market cap (in USD) based on the maximum supply of tokens |
| **MktC_C** | Market cap (in USD) based on the circulating supply of tokens |
| **COMP** | Daily price (in USD) of COMP |
| **TVL** | Value (in USD) of funds locked in the project's smart contracts |
| **Revenue** | The amount of revenue (in USD) that is distributed to COMP holders |
| **Loan vol usd** | Daily volume (in USD) of Compound loans |
| **Deposit vol usd** | Daily volume (in USD) of Compound deposits |
| **Outstanding loan** | The value (in USD) of outstanding loans in Compound |
| **Outstanding deposit** | The value (in USD) of outstanding deposits in Compound |
| **Liquidation usd** | The value (in USD) of collateral liquidated daily in Compound |
| **COMP holder** | The number of Ethereum addresses that have a non-zero balance of COMP token |
| **Active user** | Daily active users of Compound protocol |
| **Developer** | Daily active developers of Compound protocol |

Note: This table introduces the definitions of Compound-specific factors.

## A.3 Mainstream cryptocurrencies

**Table A.3: Most frequently traded cryptocurrencies in Aave**

| Panel A: Aave v1 | | | | | | | | | | | |
|---|---|---|---|---|---|---|---|---|---|---|---|
| Volume (in native units) | | | | | | Volume (in USD) | | | | | |
| Token | Deposit | Token | Borrow | Token | Repay | Token | Deposit | Token | Borrow | Token | Repay |
| USDC | 5649207710 | USDC | 1707615819 | USDC | 1681778913 | ETH | 17142656515 | USDC | 1708220210 | USDC | 1682820945 |
| USDT | 3442961720 | USDT | 1228560896 | USDT | 1203833080 | USDC | 5652310840 | USDT | 1230100662 | USDT | 1206780922 |
| DAI | 1380447528 | TUSD | 664601183.1 | TUSD | 653265579.4 | USDT | 3448781433 | ETH | 1091740319 | ETH | 1185725310 |

| | | | | | | | | | |
|---|---|---|---|---|---|---|---|---|---|
| TUSD | 1288943871 | DAI | 661581867.7 | DAI | 638854367.6 | WBTC | 2424627834 | DAI | 666143741.7 | TUSD | 654175842.9 |
| LEND | 1060450285 | BUSD | 188101907.1 | REN | 186989344.5 | LINK | 1663082341 | TUSD | 665493289.3 | LINK | 647405946.5 |
| REN | 341897508.3 | REN | 187466928.3 | BUSD | 182523388.5 | DAI | 1389148414 | LINK | 634712088.2 | DAI | 642992846.4 |
| BUSD | 299873189.2 | sUSD | 71298512.28 | sUSD | 70887710.58 | TUSD | 1290313669 | WBTC | 335130896.5 | WBTC | 362742186.9 |
| MANA | 125805716.1 | BAT | 64867038.38 | BAT | 64869642.33 | YFI | 1062782098 | YFI | 309292468.5 | YFI | 332486150.1 |
| LINK | 115076956.5 | LINK | 43478548.64 | LINK | 40683159.07 | AAVE | 316317637.8 | SNX | 224380696.9 | SNX | 233358901 |
| sUSD | 77349928.2 | MANA | 40446757 | MANA | 40089594.32 | BUSD | 299989009.1 | BUSD | 188168031.6 | BUSD | 182737973.7 |

**Panel B: Aave v2**

| Volume (in native units) | | | | | | Volume (in USD) | | | | | |
|---|---|---|---|---|---|---|---|---|---|---|---|
| Token | Deposit | Token | Borrow | Token | Repay | Token | Deposit | Token | Borrow | Token | Repay |
| DAI | 46833641257 | USDC | 30732489461 | USDC | 30012731713 | ETH | 105553652834.50 | USDC | 30763638432 | USDC | 30054106996 |
| USDC | 45408614550 | USDT | 15867496023 | USDT | 15577465824 | DAI | 46892403036 | USDT | 15891932225 | USDT | 15597023168 |
| USDT | 17480009227 | DAI | 9230162254 | DAI | 9040297983 | USDC | 45454985503 | ETH | 12443385792 | ETH | 11232475374 |
| sUSD | 2100644211 | sUSD | 1118516004 | sUSD | 1093544582 | USDT | 17055576562 | DAI | 9254134272 | DAI | 9054315431 |
| TUSD | 787748498.3 | FEI | 860498739 | FEI | 861793700.1 | WBTC | 15833679310 | WBTC | 3072630752 | WBTC | 3023202777 |
| CRV | 776997569 | CRV | 603840551.4 | TUSD | 479424658.9 | stETH | 10508631822 | sUSD | 1122832093 | sUSD | 1098054622 |
| FRAX | 528080625.2 | TUSD | 498752559.6 | CRV | 477559980.7 | LINK | 4104634079 | FEI | 858001798.5 | FEI | 860709879.4 |
| BUSD | 367213163.3 | FRAX | 425849040.9 | FRAX | 411966723.7 | AAVE | 2352221041 | CRV | 630494945.5 | CRV | 533383395.1 |
| FEI | 361360038 | BUSD | 249839947.8 | BUSD | 221899332.3 | sUSD | 2108122634 | TUSD | 499323581.3 | TUSD | 479980120.7 |
| REN | 308217616.4 | GUSD | 202274766.9 | GUSD | 191008871.9 | CRV | 857227280.8 | FRAX | 425869620.6 | FRAX | 411817433.6 |

Note: This table reports the most frequently traded cryptocurrencies in Aave v1 and v2 (from Dec 1$^{st}$, 2020 to Jan 31$^{st}$, 2023). For deposit, borrow and repay events, we calculate the volume by Jan 31$^{st}$, 2023 in native units and USD, respectively. For each category of events, we list top ten tokens.

**Table A.4: Activities related to mainstream cryptocurrencies**

| | Deposit | | Withdraw | | Borrow | | Repay | | Liquidation | |
|---|---|---|---|---|---|---|---|---|---|---|
| | Event | Users | Event | Users | Event | Users | Event | Users | Event | Users |
| **ETH** | 170892 | 18929 | 116868 | 13645 | 52808 | 4338 | 49349 | 9250 | 1211 | 477 |
| **WBTC** | 38798 | 10117 | 25950 | 7340 | 10420 | 2618 | 9882 | 2357 | 199 | 102 |
| **USDC** | 79350 | 23660 | 78488 | 18587 | 125639 | 35401 | 80300 | 17214 | 7758 | 2770 |
| **USDT** | 36868 | 13035 | 36823 | 10021 | 60178 | 12232 | 44265 | 10754 | 5101 | 1798 |
| **DAI** | 52579 | 19989 | 44558 | 14475 | 51997 | 10712 | 42435 | 9027 | 4197 | 1423 |
| **TUSD** | 4220 | 1557 | 6227 | 1271 | 13350 | 2641 | 8460 | 2432 | 1250 | 341 |
| **sUSD** | 8485 | 3263 | 7292 | 2831 | 5402 | 1670 | 4730 | 1526 | 441 | 155 |
| **BUSD** | 3334 | 2340 | 3432 | 1712 | 7233 | 2345 | 4704 | 1813 | 823 | 233 |
| **FEI** | 492 | 150 | 483 | 134 | 648 | 224 | 638 | 203 | 33 | 16 |
| **FRAX** | 697 | 374 | 494 | 261 | 1060 | 466 | 985 | 377 | 102 | 32 |
| **AAVE** | 13361 | 5659 | 10108 | 4052 | 424 | 30 | 425 | 30 | 0 | 0 |
| **CRV** | 12637 | 7116 | 6815 | 3918 | 3373 | 1104 | 4283 | 950 | 451 | 32 |
| **SNX** | 16432 | 8390 | 11474 | 6239 | 2712 | 966 | 2624 | 921 | 136 | 53 |
| **YFI** | 11389 | 3897 | 9298 | 3195 | 2145 | 759 | 2145 | 734 | 93 | 23 |
| **LINK** | 49855 | 10839 | 31220 | 8383 | 5921 | 1951 | 5720 | 1857 | 435 | 72 |
| **stETH** | 31057 | 22125 | 8547 | 3968 | 171 | 7 | 171 | 7 | 0 | 0 |

Note: This table reports descriptive statistics of the datasets for Aave V1 and V2 lending pool (from Dec 16$^{th}$, 2019 to Jan 31$^{st}$, 2023). The datasets include five main categories of events, i.e., deposit, withdraw, borrow, repay and liquidation. For each cryptocurrency and each category of events, the number of events and unique participants in the events are presented.

**Table A.5: Descriptive statistics of loans and deposits of mainstream cryptocurrencies traded in Aave**

| Panel A: Loan details | | | | | |
|---|---|---|---|---|---|
| | **Mean** | **Median** | **Maximum** | **Minimum** | **Std** |
| **Borrower** | 170.25 | 151 | 2190 | 3 | 134.19 |
| **Loan vol usd** | 72683824.09 | 42963471.59 | 1311529749.84 | 412.73 | 110689522.00 |
| **Loan cnt** | 241.35 | 210 | 2248 | 8 | 177.35 |
| **New borrower** | 47.24 | 33 | 2112 | 1 | 93.00 |
| **New loan vol usd** | 10755476.93 | 2079904.68 | 450786027.67 | 4.92 | 34996821.37 |
| **New loan cnt** | 55.68 | 40 | 2140 | 1 | 96.18 |
| **Avg loan usd** | 291362.72 | 176986.09 | 6841263.49 | 29.48 | 406080.79 |
| **Repeat borrower** | 123.01 | 101 | 529 | 0 | 87.79 |
| **Repeat loan vol usd** | 61928347.16 | 37229370.52 | 1292426346.84 | 0.00 | 93667368.13 |
| **Repeat loan cnt** | 185.67 | 155 | 881 | 0 | 132.97 |
| **Panel B: Deposit details** | | | | | |
| | **Mean** | **Median** | **Maximum** | **Minimum** | **Std** |
| **Depositor** | 202.12 | 176 | 2287 | 1 | 143.78 |
| **Deposit vol usd** | 240907261.60 | 130627962.49 | 4350974988.98 | 3.01 | 376351664.08 |
| **Deposit cnt** | 414.57 | 340.50 | 2590 | 1 | 294.09 |
| **New depositor** | 82.63 | 63 | 2112 | 1 | 100.26 |
| **New deposit vol usd** | 32188435.46 | 6979968.68 | 3237955250.28 | 3.01 | 134973004.61 |
| **New deposit cnt** | 102.96 | 78.50 | 2182 | 1 | 111.87 |
| **Avg deposit usd** | 622002.77 | 320228.07 | 13253127.69 | 3.01 | 1154670.71 |
| **Repeat depositor** | 119.50 | 101 | 545 | 0 | 72.04 |
| **Repeat deposit vol usd** | 208718826.14 | 115055674.10 | 4234136055.94 | 0 | 325247426.38 |
| **Repeat deposit cnt** | 311.61 | 256 | 1843 | 0 | 228.95 |

Note: This table reports details of loans and deposits of mainstream cryptocurrencies traded in Aave v1 and v2 (from Dec 1st, 2020 to Jan 31st, 2023).

**Figure A.3: Depositors of mainstream cryptocurrencies in Aave protocol (Dec 16, 2019 – Jan 31, 2023)**

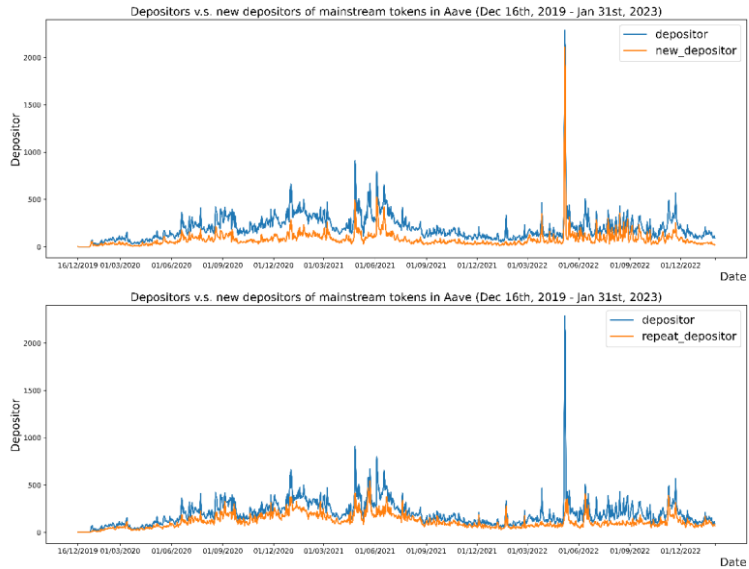

Note: This figure illustrates the number of depositors of mainstream cryptocurrencies traded in Aave protocol, and we also show the number of new depositors and repeat depositors.

**Figure A.4: Borrowers of mainstream cryptocurrencies in Aave protocol (Dec 16, 2019 – Jan 31, 2023)**

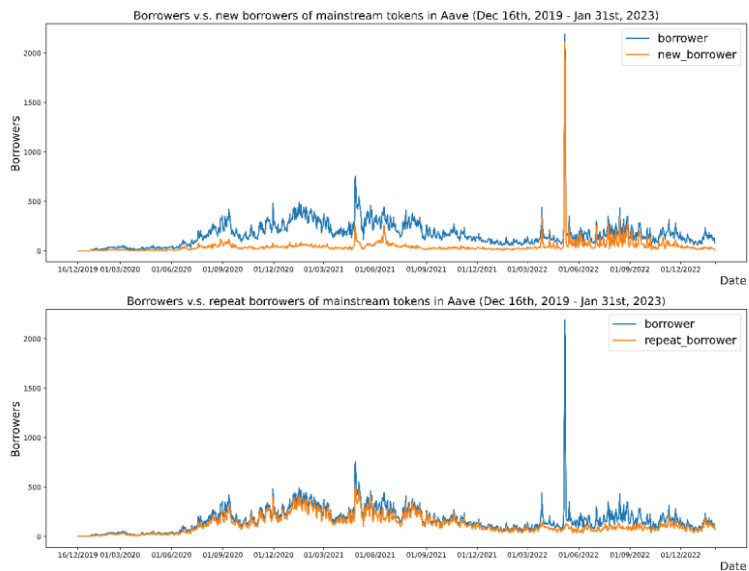

Note: This figure illustrates the number of borrowers of mainstream cryptocurrencies traded in Aave protocol, and we also show the number of new borrowers and repeat borrowers.

**Figure A.5: Deposit volume of mainstream cryptocurrencies in Aave protocol (Dec 16, 2019 – Jan 31, 2023)**

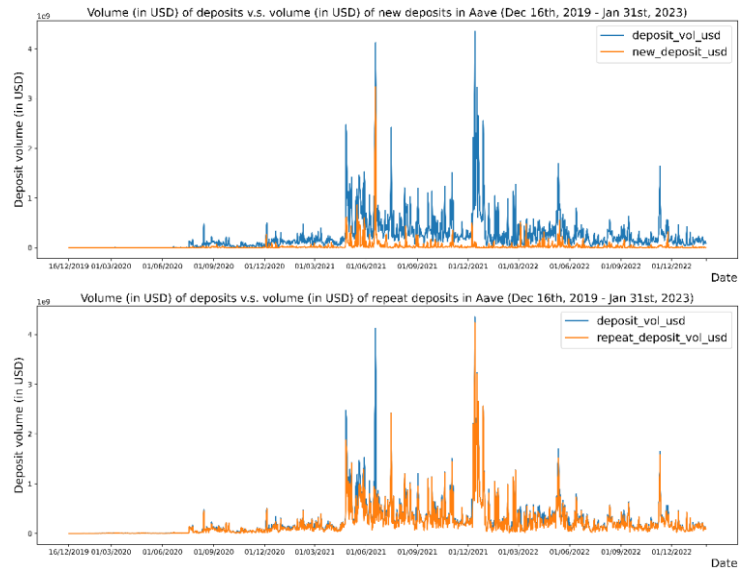

Note: This figure illustrates the volume (in USD) of deposits of mainstream cryptocurrencies traded in Aave protocol, and we also show the volumes (in USD) of deposits from new depositors and repeat depositors.

**Figure A.6: Loan volume of mainstream cryptocurrencies in Aave protocol (Dec 16, 2019 – Jan 31, 2023)**

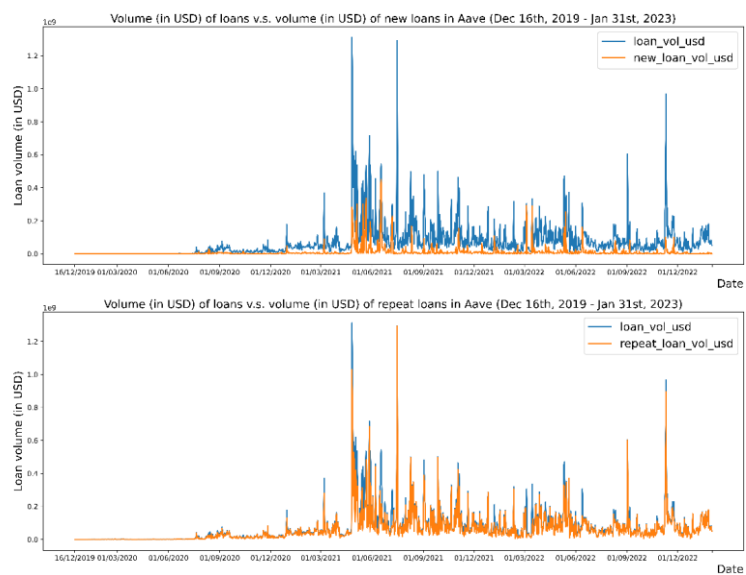

Note: This figure illustrates the volume (in USD) of loans of mainstream cryptocurrencies traded in Aave protocol, and we also show the volumes (in USD) of loans from new borrowers and repeat borrowers.

**Figure A.7: The proportion of repeat deposits and repeat loans of mainstream cryptocurrencies (Dec 16, 2019 – Jan 31, 2023)**

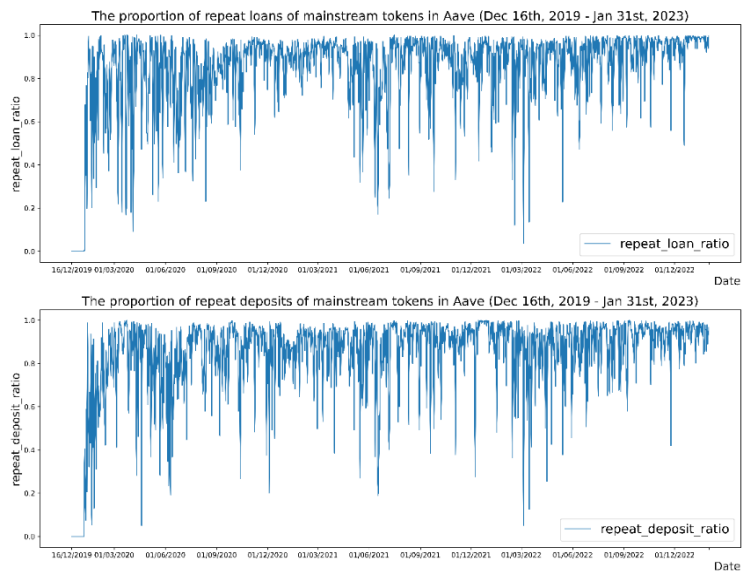

Note: This figure illustrates the proportion of repeat loans and deposits in Aave protocol, respectively.

**Footnotes**

[i] https://compound.finance/

[ii] https://aave.com

[iii] The address of 'Aave: LendingPool V1' is 0x398eC7346DcD622eDc5ae82352F02bE94C62d119

[iv] The address of 'Aave: LendingPool V2' is 0x7d2768dE32b0b80b7a3454c06BdAc94A69DDc7A9

[v] Figures A.1 and A.2 shows the numbers of daily new depositors and daily new borrowers, respectively.

[vi] Online appendix OA.1 explains how we selected independent variables included in regression models in this section.

[vii] More details about Aave V2 can be found on the official website:
https://docs.aave.com/developers/v/2.0/change-from-v1-to-v2

[viii] We do not choose Compound V2 because it was launched on May 23rd, 2019, which is earlier than the start of our datasets. More details about Compound V3 can be found on the official website: https://docs.compound.finance/